\documentclass[apj,appendixfloats,numberedappendix]{emulateapj}

\usepackage{graphicx}
\usepackage{color}

\usepackage[breaklinks,colorlinks,citecolor=black]{hyperref}

\shorttitle{TDFs are due to TDEs}
\shortauthors{van Velzen} 

\begin{document}

\title{On the mass and luminosity functions of tidal disruption flares: \\ rate suppression due to black hole event horizons}

\author{S. van Velzen\altaffilmark{1}}

\affil{Department of Physics \& Astronomy, The Johns Hopkins University, Baltimore, MD 21218}
\affil{Center for Cosmology and Particle Physics, New York University, NY 10003}

\altaffiltext{1}{Hubble Fellow, \texttt{sjoert@jhu.edu}}

\begin{abstract}
The tidal disruption of a star by a massive black hole is expected to yield a luminous flare of thermal emission. About two dozen of these stellar tidal disruption flares (TDFs) may have been detected in optical transient surveys. However, explaining the observed properties of these events within the tidal disruption paradigm is not yet possible. This theoretical ambiguity has led some authors to suggest that optical TDFs are due to a different process, such as a nuclear supernova or accretion disk instabilities. Here we present a test of a fundamental prediction of the tidal disruption event scenario: a suppression of the flare rate due to the direct capture of stars by the black hole. Using a recently compiled sample of candidate TDFs with black hole mass measurements, plus a careful treatment of selection effects in this flux-limited sample, we confirm that the dearth of observed TDFs from high-mass black holes is statistically significant. All the TDF impostor models we consider fail to explain the observed mass function; the only scenario that fits the data is a suppression of the rate due to direct captures. We find that this suppression can explain the low volumetric rate of the luminous TDF candidate ASASSN-15lh, thus supporting the hypothesis that this flare belongs to the TDF family. Our work is the first to present the optical TDF luminosity function. A steep power law is required to explain the observed rest-frame $g$-band luminosity, $dN/dL_{g} \propto L_{g}^{-2.5}$. The mean event rate of the flares in our sample is $\approx 1 \times10^{-4}$~{galaxy}$^{-1}$\,yr$^{-1}$, consistent with the theoretically expected tidal disruption rate. 
\end{abstract}

\maketitle 

\section{Introduction}
A stellar tidal disruption event (TDE) happens when a star passes within the Roche radius of a massive black hole. As the streams of stellar debris circularize and are accreted by the black hole, we can expect a luminous flare of thermal emission \citep{Rees88}. The first (candidates) of these stellar tidal disruption flares (TDFs) were discovered with X-ray surveys \citep{Bade96,KomossaBade99}, followed by UV-selected flares \citep{Gezari06,Gezari09}. The detection of new TDFs is currently dominated by optical surveys \citep{vanVelzen10,Gezari12,Arcavi14,Holoien14}. About two new candidates are found each year, and this number is certain to increase in the near future as larger optical surveys (ZTF, LSST) become operational.

Stellar tidal disruption flares can be considered a multipurpose tool for extragalactic astrophysics. First of all, they can be used as probes to signal the existence of black holes in quiescent galaxies, potentially including intermediate-mass black holes in dwarf galaxies \citep{Maksym13,MacLeod14}. The dynamics of the TDE itself are also interesting; the rapid increase of the fallback rate of the stellar debris (from super- to sub-Eddington in a few years) presents a new domain to test our understanding of accretion physics. In particular, the detection of radio emission following TDFs \citep{Zauderer11,Bloom11,Levan11,vanVelzen16,Alexander17} can shed new light on the conditions required for the launch of relativistic jets. 

While TDFs appear to be promising tools, their optical emission has proven to be difficult to model self-consistently. The observed temperature is too low to originate from a compact accretion disk near the pericenter of the stellar orbit \citep{Strubbe11,Lodato11,vanVelzen10}. Reprocessing of high-energy photons from the compact accretion disk by material at larger radii \citep{Guillochon14,Metzger16} and emission due to shocks caused by intersecting debris streams \citep{Piran15b,Krolik16} have both been proposed as explanations for optical TDF emission. Neither explanation is completely satisfactory. The intersecting-stream scenario suffers from a ``missing energy problem'' \citep{Piran15b}, whereas the reprocessing scenario appears at odds with the (tentative) discovery  that the X-ray emission lags the UV emission \citep{Pasham17}.

Contrary to the optical emission, the X-ray properties of TDF candidates generally fit within the canonical TDE picture of \citet{Rees88}; see \citet[][]{Komossa15,Auchettl16} for reviews. Since very few X-ray selected TDFs have received optical follow-up near peak \citep{Saxton17}, it could be possible that X-ray-selected and optically selected TDF candidates are a separate class of tidal disruptions \citep{Dai15}. Some authors have gone one step further and proposed that most optically selected TDF candidates are, in fact, not due to the tidal disruption of a star. Supernovae (SNe) in AGN accretion disks \citep{Saxton16}, collisions of stars on bound orbits around the black hole \citep{Metzger17}, black hole accretion disk instabilities \citep{Saxton16}, or flares from accreting supermassive black hole binaries with subparsec separations \citep{Tanaka13} have been proposed as potential TDF impostors.

The hypothesis that flares from active galactic nuclei (AGN) are TDF impostors perhaps has the most observational support because relatively rapid changes of AGN luminosity and spectral type have been observed \citep{Storchi-Berg93,Shappee14,LaMassa15,Gezari17}. While the observed properties of these changing-look AGN are different from candidate TDFs \citep[][]{Ruan16,MacLeod16}, their existence demonstrates that the luminosity of AGN can increase beyond what is expected from the power spectrum that accurately describes the light curves of mundane AGN \citep{MacLeod12,Graham17}. 

The mechanism that triggers the accretion rate increase in changing-look AGN could be similar to the thermal-viscous instability \citep{Meyer81,Smak83} that is known to operate in disks around stellar-mass compact objects that accrete from a donor star \citep{van-Paradijs96,Lasota01}. For black holes with a mass of $10^{8}\,M_{\odot}$, a similar mechanism \citep{Mineshige90,Siemiginowsk97,Janiuk02} could lead to outbursts with a recurrence time of $10^{3-6}\,~{\rm yr}$ \citep{Hatziminaogl01,Czerny09}. However, \citet*{Hameury09} argue that due to the high Mach number in AGN accretion disks, this quiescent time could be much shorter, and therefore large outbursts due to limit-cycle oscillations may not be expected for AGN.  

The tidal disruption of a star can also occur in a Seyfert galaxy \citep[e.g.,][]{Drake11,Blanchard17}---perhaps such events could explain some of the changing-look AGN \citep{Merloni15,Wyrzykowski17}. Identifying observations that can discriminate between an AGN disk instability triggered by a star passing through the disk and a TDE inside a galaxy that also harbors an AGN is very challenging; these events are a mix of two phenomena that are not yet completely understood even if they occur in isolation. Fortunately, we can largely avoid this problem if we consider only TDF candidates from galaxies that are not classified as AGN before the optical outburst. 

One signature is unique to a TDE and can thus be used to rule out most---or perhaps all---TDF impostor scenarios. A TDE requires that the star passes within the tidal radius, 
\begin{equation}\label{eq:Rt}
	R_{t} = R_{\rm star} \left(\frac{M_{\bullet}}{M_{\rm star}} \right)^{1/3} \sim 25 R_{s} \left(\frac{M_{\bullet}}{10^{6}M_{\odot}}\right)^{{-2/3}} \quad. 
\end{equation}
Here we expressed the tidal radius for a solar-type star in units of the Schwarzschild radius ($R_{s}\equiv 2GM_{\bullet}/c^{2}$). For a black hole mass of $M_{\bullet} \approx 10^{8} M_{\odot}$ the tidal radius of a solar-type star is equal to the Schwarzschild radius, and the results of the disruption will not be visible to an observer outside the black hole event horizon \citep{Hills75}. This critical black hole mass is sometimes referred to as the Hills mass.

Repeating the calculation behind Eq.~\ref{eq:Rt} within in the framework of General Relativity, one finds that the Hills mass increases with black hole spin \citep{Kesden12b}. For spinning black holes, the outcome of the disruption also depends on the orientation between the black hole angular momentum vector and the star's orbital plane; after averaging over all inclinations, \citet{Kesden12b} finds that the effect of the black hole horizon yields a superexponential suppression of the TDF rate for $M_{\bullet}>M_{\rm Hills}$. 

The goal of this work is to use the expected suppression in the flare rate due to a black hole horizon to demonstrate that optically selected TDF candidates are indeed due to the tidal disruption of a star. 

The outline of this paper is as follows. We will first present our compilation of TDF candidates in Sec.~\ref{sec:sample}. Next, we compute the luminosity function (LF) and host galaxy mass function for this sample (Sec.~\ref{sec:LF}). We then use forward modeling to reproduce the observed distribution of black hole mass and galaxy mass (Sec.~\ref{sec:mock}). We discuss the implications of this result (Sec.~\ref{sec:discussion}) and close with a list of conclusions (Sec.~\ref{sec:conclusion}). We adopt a flat cosmology with $\Omega_{\Lambda}=0.7$ and $H_{0}=70\,{\rm km}{\rm s}^{-1}\,{\rm Mpc}^{-1}$. All magnitudes are in the AB \citep{oke74} system. 

\subsection{TDE or TDF?}
In the literature, both tidal disruption event (TDE) and tidal disruption flare (TDF) are used to label transients due to stellar disruptions. In this work, we use TDE to refer to the general concept of the disruption of a star by a black hole, while TDF is used only for the electromagnetic result of this disruption. This distinction is subtle, yet useful, since not every TDE may lead to a TDF \citep[e.g.,][]{Guillochon15}.

\begin{deluxetable*}{l l l c c c c c c}
\tablewidth{0pt}

\tablecolumns{9}
\tablecaption{Sample of 17 candidate TDFs. }
\tablehead{name & R.A. & Decl. & $m_{\rm max}$ & $L_{g}$ &  $T$ & $L_{\rm bb}$ & $z$ & $z_{\rm max}$ \\
          			 & (J2000) & (J2000) &  & ($\log_{10} {\rm erg}\,{\rm s}^{-1})$ & ($\times 10^{4}$ K) & ($\log_{10} {\rm erg}\,{\rm s}^{-1})$ &  & \\}
\startdata
GALEX-D1-9      & 02:25:16.96& $-$04:32:59.1 & 22.4 (NUV) &  $42.3$  & 5.6 & 44.1 & 0.326  & 0.554\\
GALEX-D3-13     & 14:19:29.78& $+$52:52:06.3 & 22.2 (NUV)$^*$ &  $42.7$  & 4.9 & 44.3 & 0.370  & 0.821\\
GALEX-D23H-1    & 23:31:59.53& $+$00:17:14.5 & 20.9 (NUV) &  $42.3$  & 4.9 & 43.9 & 0.185  & 0.517\\
SDSS-TDE1       & 23:42:01.40& $+$01:06:29.2 & 21.0 ($r$)$^*$ &  $42.7$  & 2.4 & 43.5 & 0.136  & 0.174\\
SDSS-TDE2       & 23:23:48.61& $-$01:08:10.2 & 20.3 ($r$)$^*$ &  $43.5$  & 1.8 & 44.0 & 0.256  & 0.469\\
PS1-10jh        & 16:09:28.27& $+$53:40:23.9 & 19.8 ($g$) &  $43.2$  & 2.9 & 44.2 & 0.170  & 0.409\\
PS1-11af        & 09:57:26.81& $+$03:14:00.9 & 21.4 ($g$) &  $43.3$  & 1.9 & 43.9 & 0.405  & 0.426\\
PTF-09ge        & 14:57:03.18& $+$49:36:40.9 & 17.7 ($r$) &  $43.4$  & 2.2 & 44.1 & 0.064  & 0.155\\
PTF-09axc       & 14:53:13.07& $+$22:14:32.2 & 19.1 ($r$) &  $43.2$  & 1.2 & 43.5 & 0.115  & 0.138\\
PTF-09djl       & 16:33:55.97& $+$30:14:16.6 & 19.6 ($r$) &  $43.5$  & 2.6 & 44.4 & 0.184  & 0.179\\
ASASSN-14ae     & 11:08:40.11& $+$34:05:52.2 & 17.0 ($g$)$^*$ &  $43.2$  & 2.1 & 43.9 & 0.044  & 0.051\\
ASASSN-14li     & 12:48:15.23& $+$17:46:26.4 & 16.8 ($g$)$^*$ &  $42.6$  & 3.5 & 43.8 & 0.021  & 0.025\\
ASASSN-15oi     & 20:39:09.14& $-$30:45:20.6 & 16.2 ($g$)$^*$ &  $43.6$  & 2.5 & 44.4 & 0.048  & 0.082\\
ASASSN-15lh     & 22:02:15.44& $-$61:39:34.5 & 16.5 ($g$) &  $44.8$  & 2.5 & 45.6 & 0.233  & 0.347\\
iPTF-15af       & 08:48:28.13& $+$22:03:33.4 & --  &  --  & -- & -- & 0.079  & 0.101\\
iPTF-16axa      & 17:03:34.34& $+$30:35:36.6 & 18.5 ($r$)$^*$ &  $43.5$  & 3.0 & 44.5 & 0.108  & 0.135\\
iPTF-16fnl      & 00:29:57.05& $+$32:53:37.2 & 17.4 ($r$) &  $42.3$  & 3.0 & 43.4 & 0.016  & 0.034\\

\enddata
\tablecomments{The third column, $m_{\rm max}$,  lists the maximum observed apparent magnitude and the relevant filter for each survey; an asterisk indicates that the peak of the light curve was not resolved. The forth and fifth columns give the rest-frame $g$-band luminosity, as computed using the observed blackbody temperature ($T$) to make the k-correction. The last column lists the maximum redshift where this flare could have been detected given the survey effective flux limit.}
\label{tab:TDFs}
\end{deluxetable*}

\begin{deluxetable*}{l c c c c c c c c c c}
\tablewidth{0pt}
\tablecolumns{10}
\tablecaption{Properties of the TDF host galaxies. }
\tablehead{name & $m_{g}$ & $m_{r}$ & $m_{K}$ & $M_{r}$ & $M_{g}$ & gal. mass & $\sigma$ & $M_{\bullet}$ & $z_{\rm max, \sigma}$ \\
          &   & &  & &  & $(\log_{10} M_{\odot})$ & (km\,s$^{-1})$ & $(\log_{10} M_{\odot})$ &  \\}
\startdata
GALEX-D1-9      & $22.01\pm0.12$ & $20.92\pm0.05$ & $19.29\pm0.01$ & $-20.5$ & $-20.0$ & 10.3 & $       -   $ & $    -   $ & 0.324 \\
GALEX-D3-13     & $21.97\pm0.07$ & $20.45\pm0.03$ & $18.40\pm0.01$ & $-21.5$ & $-20.8$ & 10.7 & $  133 \pm 6$ & $7.4\pm 0.4$ & 0.375 \\
GALEX-D23H-1    & $20.09\pm0.03$ & $19.24\pm0.02$ & $17.81\pm0.07$ & $-20.7$ & $-20.1$ & 10.3 & $   77 \pm18$ & $6.4\pm 0.6$ & 0.395 \\
SDSS-TDE1       & $20.27\pm0.02$ & $19.24\pm0.02$ & $17.93\pm0.07$ & $-19.9$ & $-19.2$ & 10.1 & $  126 \pm 7$ & $7.3\pm 0.4$ & 0.275 \\
SDSS-TDE2       & $20.79\pm0.05$ & $19.50\pm0.02$ & $18.02\pm0.07$ & $-21.3$ & $-20.6$ & 10.6 & $       -   $ & $    -   $ & 0.410 \\
PS1-10jh        & $21.91\pm0.08$ & $21.05\pm0.05$ & $19.88\pm0.02$ & $-18.6$ & $-18.1$ &  9.5 & $   65 \pm 3$ & $6.1\pm 0.4$ & 0.170 \\
PS1-11af        & $22.89\pm0.23$ & $21.35\pm0.09$ & $19.59\pm0.23$ & $-20.7$ & $-20.1$ & 10.1 & $       -   $ & $    -   $ & 0.284 \\
PTF-09ge        & $17.91\pm0.01$ & $17.13\pm0.01$ & $16.71\pm0.10$ & $-20.2$ & $-19.5$ & 10.1 & $   72 \pm 6$ & $6.2\pm 0.4$ & 0.330 \\
PTF-09axc       & $18.66\pm0.01$ & $18.04\pm0.01$ & $17.60\pm0.15$ & $-20.7$ & $-20.2$ & 10.0 & $   60 \pm 4$ & $5.9\pm 0.4$ & 0.411 \\
PTF-09djl       & $20.57\pm0.03$ & $19.70\pm0.02$ & $18.54\pm0.11$ & $-20.2$ & $-19.6$ & 10.1 & $   64 \pm 7$ & $6.0\pm 0.5$ & 0.323 \\
ASASSN-14ae     & $17.27\pm0.01$ & $16.64\pm0.01$ & $16.23\pm0.08$ & $-19.9$ & $-19.2$ &  9.8 & $   53 \pm 2$ & $5.7\pm 0.4$ & 0.291 \\
ASASSN-14li     & $15.98\pm0.00$ & $15.47\pm0.00$ & $14.98\pm0.04$ & $-19.3$ & $-18.8$ &  9.6 & $   78 \pm 2$ & $6.4\pm 0.4$ & 0.249 \\
ASASSN-15oi     & $17.44\pm0.01$ & $16.79\pm0.02$ & $15.90\pm0.07$ & $-19.9$ & $-19.3$ &  9.9 & $       -   $ & $    -   $ & 0.301 \\
ASASSN-15lh     & $19.59\pm0.10$ & $18.32\pm0.10$ & $17.01\pm0.12$ & $-22.2$ & $-21.4$ & 10.8 & $  225 \pm15$ & $8.3\pm 0.4$ & 0.571 \\
iPTF-15af       & $18.36\pm0.01$ & $17.49\pm0.01$ & $17.02\pm0.11$ & $-20.4$ & $-19.6$ & 10.2 & $  106 \pm 2$ & $7.0\pm 0.4$ & 0.341 \\
iPTF-16axa      & $19.33\pm0.02$ & $18.46\pm0.01$ & $17.92\pm0.17$ & $-20.1$ & $-19.4$ & 10.1 & $   82 \pm 3$ & $6.5\pm 0.4$ & 0.314 \\
iPTF-16fnl      & $15.22\pm0.00$ & $14.72\pm0.00$ & $14.18\pm0.07$ & $-19.5$ & $-19.0$ &  9.8 & $   55 \pm 2$ & $5.8\pm 0.4$ & 0.277 \\
\enddata
\tablecomments{The apparent magnitudes ($m_{r}$, $m_{g}$, and $m_{K}$) are corrected for Galactic extinction using the \citet{Schlafly11} extinction maps. Uncertainties on the apparent magnitudes include only the statistical uncertainty. The absolute magnitudes are computed in the rest frame of the host galaxy.  
The total stellar mass of the galaxy is estimated from the broadband ({\it ugrizJHK}) photometry. The velocity dispersion ($\sigma$) measurements are from the sample of \citet{Wevers17} (with the exception of ASASSN-15lh). The last column lists the maximum redshift for inclusion of the galaxy in the \citet{Wevers17} sample.}
\label{tab:hosts}
\end{deluxetable*}

%
%

%
\section{Observed TDFs}\label{sec:sample}
%
We will restrict our sample of candidate TDFs to nuclear flares found in optical/UV imaging surveys. The first motivation for this choice is the fact that candidate TDFs from optical imaging surveys show a number of common properties (discussed below), which justifies treating these flares as one class in our analysis. Further motivation for restricting to optical surveys is that other methods to find TDFs, X-ray surveys, or extreme coronal line emitters in spectroscopic galaxy samples \citep{Komossa08} require more assumptions to estimate the event rate. For these surveys, the cadence is (much) lower than the duration of the flare and estimating the volumetric flare rate requires a light-curve model (or one could measure the snapshot rate; see \citealt*{Auchettl17}).  

We exclude flares found in galaxies that can be classified as a broad-line AGN or Seyfert, but we include flares from LINERs \citep{heckman80}, leaving 17 sources (Table~\ref{tab:TDFs}). { Our selection requirements and final sample are similar to the candidate TDF samples presented recently \citep{Hung17,Wevers17}. The only difference is that this earlier work used stricter requirements on either the TDF light-curve sampling (to be able to measure the decay rate) or the host brightness and Declination (to allow efficient spectroscopic follow-up).}

These 17 flares share a number of properties: a high blackbody temperature ($T=[1-3] \times 10^{4}$~K) and nearly constant colors (e.g., \citealt{Hung17}, Fig.~11). Also the optival/UV light curves of the TDFs in our sample are all consistent with a power-law decay rate; in all cases where monitoring observations cover more than one year of the light curve, an exponential decay rate can be ruled out \citep{vanVelzen11,Arcavi14,Gezari15,vanVelzen16,Brown16,Brown16b}. 

In Table~\ref{tab:TDFs} we summarize the observed properties of the flares in our sample. To measure the rest-frame $g$-band luminosity ($L_{g}$) we used the observed blackbody luminosity to compute the k-correction \citep{Hogg02}.

\begin{deluxetable}{l c c c c r}
\tablewidth{0pt}
\tablecolumns{6}
\tablecaption{Survey properties. }
\tablehead{Survey & $N_{\rm TDF}$ & $m_{\rm lim}$ & band & $z_{\rm max^{*}}$ & {$(A\times \tau)^{*}$}\\
					  & & & & & (deg$^{2}$~yr)\\}
\startdata
GALEX      & 3 & 23.0 & NUV   & 0.393 &       17 \\
SDSS       & 2 & 21.5 & r     & 0.140 &      202 \\
PS1        & 2 & 21.5 & g     & 0.168 &      120 \\
PTF        & 3 & 19.5 & r     & 0.054 &     3000 \\
iPTF       & 3 & 19.5 & r     & 0.054 &     5032 \\
ASAS-SN    & 4 & 17.3 & g     & 0.023 &    82637 \\
\enddata
\tablecomments{The second column lists the number of candidate TDFs from each survey. The typical maximum redshift for each survey ($z_{\rm max*}$) follows from the requirement that a flare with a peak luminosity of $L_{g}=10^{42.5}\,{\rm erg}\,{\rm s}^{-1}$ is detectable above the effective flux limit ($m_{\rm lim}$). From this redshift we can compute (Eq.~\ref{eq:rate}) the effective area times the survey duration.}
\label{tab:surveys}
\end{deluxetable}

Measurements of the velocity dispersion of the host galaxy have been obtained by \citet{Wevers17} for 12 of the 17 TDFs in our sample. This sample includes only sources at Declination $>0$ and is complete for a host galaxy flux limit of $m_{g}<22$ and $m_{r}<21$.  

The stellar mass of the TDF host galaxies is estimated from broadband optical to near-IR photometry using \verb kcorrect   \citep{blanton07}. The same software and wavelength range was used to estimate the mass of galaxies that are input for our synthetic sample of potential TDF host galaxies (see Sec.~\ref{sec:mock}). We use the SDSS \citep{york02} Petrosian \citep{blanton01,stoughton02} magnitudes (the treatment of the few flares outside the SDSS footprint is discussed below). The IR flux in the {\it J, H}, and {\it K} bands is measured from images of 2MASS \citep{Skrutskie97,Jarrett00} using a circular aperture with a radius equal to the 90\% light radius in $r$-band. When available, we substitute the 2MASS images with UKIDSS images \citep{Lawrence07,Hambly08} since the latter provide a better signal-to-noise ratio. 

The TDFs in our sample were discovered by different surveys, each with their own selection function and detection efficiency. Since the detection efficiency of most of these surveys is unknown, we cannot use our sample to obtain an {\it absolute} measurement of the event rate or luminosity function. However, since each survey discovers events from the same parent distribution, we can use the detected number of TDF candidates in each survey to compare the selection efficiencies and thus obtain the {\it relative} luminosity function. 

The detected number of flares in a given imaging survey is a linear function of the survey area, efficiency, and survey duration and a nonlinear function of the survey effective flux limit ($m_{\rm lim}$). Because multiple detections or spectroscopic follow-up observations of the flare are often required to identify a transient as a candidate TDF, the effective flux limit is typically larger than the single-epoch detection limit of the survey images. We estimate $m_{\rm lim}$ from the observed distribution of the flare's apparent magnitude near peak in each survey. 

The effective flux limit can be used to compute the maximum redshift, $z_{\rm max}$, for the detection and identification of a flare with a given peak luminosity ($L_{g}$) and temperature. 
For each of the five surveys in our sample we can thus compute the volume in which a typical TDF can be detected. Here we define a typical TDF as a flare with a peak luminosity of $L_{g}^{*} = 10^{42.5}$~erg~s$^{-1}$ and temperature $T^{*} = 2.5 \times 10^{4}$~K. The number of detected TDFs in each survey can now be estimated as {
\begin{equation}\label{eq:rate}
N_{\rm TDF,\,detected} \approx \dot{N} \, V(z_{\rm max*}) ~ A_{\rm survey} \times \tau_{\rm survey} \quad.
\end{equation}
Here $A_{\rm survey}\times \tau_{\rm survey}$ is used to label the product of the effective survey area and duration. $V(z_{\rm max*})$ denotes the comoving volume (per solid angle) corresponding to $z_{\rm max}$}. Since the flares in our sample span a relatively narrow redshift range, we may assume that each survey is sensitive to the same event rate ($\dot{N}$), and thus the number of TDFs found in each survey can be used to estimate $A_{\rm survey}\times \tau_{\rm survey}$. In Table~\ref{tab:surveys} we summarize the results of this exercise. To set the normalization of $A_{\rm survey}$, we adopted a mean rate of $\dot{N} = 5 \times 10^{-7} {\rm Mpc}^{-3} {\rm yr}^{-1}$, which was chosen to match the volumetric rate based on SDSS and ASAS-SN data \citep{vanVelzen14,Holoien16}. 

Comparing the value of $A_{\rm survey}\times \tau_{\rm survey}$ obtained from Eq.~\ref{eq:rate} with the  area and duration of each survey yields an estimate of the mean detection efficiency. For example, our estimate of {$(A \times \tau)^{*}$} for the ASAS-SN and GALEX TDF searches is similar to the total area and duration of these surveys, implying a high efficiency for detecting and identifying TDFs. 

\subsection{Input Surveys}
In the following subsections we briefly discuss the surveys that provided the input for our compilation of TDF candidates.

\subsubsection{GALEX}
Three flares in our sample were found by searching for transients in GALEX \citep{Martin05} multi-epoch imaging in the near-UV (NUV) and far-UV (FUV) bands: GALEX-D3-13 \citep{Gezari06} GALEX-D1-9 \citep{Gezari08}, and GALEX-D23H-1 \citep{Gezari09}. The search was conducted using $\approx 5$ yr of GALEX observations of four extragalactic fields (each covering $\sim 1$~deg$^{2}$ of the sky). 
Candidate TDFs were selected as transient UV sources from inactive galaxies. Active galaxies were identified using the large body of archival spectroscopic observations that are available for these fields, complemented by spectroscopic follow-up observations where necessary \citep{Gezari08}. For this survey, we adopt an effective flux limit of $m<23$ in the NUV band. The source D3-13 is located in the CANDLES \citep{Grogin11,Koekemoer11} footprint, and we use the WIRCam data \citep{Bielby12} cataloged by \citet{Stefanon17} to measure the near-IR flux of its host galaxy.  

\subsubsection{SDSS Stripe~82}
Two flares in our sample, TDE1 and TDE2 \citep{vanVelzen10}, were found by searching for transients in SDSS Stripe~82 multi-epoch imaging data \citep{frieman08,Abazajian09}, covering about 300~deg$^{2}$. This search used the SDSS $u$, $g$, and $r$ filters and selected nuclear transients from inactive galaxies. Active galaxies were identified using SDSS spectra, colors, and optical variability \citep{vanVelzen10}.  For this survey, we adopt an effective flux limit of 21.5 in the $r$ band. 

\subsubsection{Pan-STARRS}
Two flares in our sample originate from the Pan-STARRS \citep{Chambers07,Chambers16} Medium Deep (PS1 MD) fields: PS1-10jh \citep{Gezari12} and PS1-11af \citep{Chornock14}. These two candidate TDFs did not originate from a single search, but since the PS1 MD fields all have a similar single-epoch flux limit, we will treat them as originating from one survey. For this survey, we adopt an effective flux limit of 21.5 in the $g$ band. 

\subsubsection{PTF}\label{sec:PTF}
Three flares in our sample originate from the analysis of \citet{Arcavi14} using the Palomar Transient Factory \citep[PTF;][]{Law09}: PTF-09ge, PTF-09djl, and PTF-09axc. These TDF candidates were obtained by selecting nuclear transients from PTF imaging data that have received spectroscopic follow-up observations and have peak $R$-band luminosity in the range $-21<M_{r}<-19$. For this survey, we adopt an effective flux limit of 19.5 in the $r$ band. 
Since the PTF search has a restriction on the TDF luminosity, our method will underestimate the effective area ($A_{\rm survey}$ via Eq.~\ref{eq:rate}) if the rate decreases with increasing flare luminosity. The TDF LF (discussed in the next section) indeed has a negative slope, and we modify our estimate of $A_{\rm survey}$ (listed in Table~\ref{tab:surveys}) to take this into account. 

Besides the three sources presented by \citet{Arcavi14}, the PTF survey has yielded one more TDF candidate: PTF10iya \citep{Cenko12}. We exclude this source from our compilation since the WISE \citep{Wright10} colors provide strong evidence for a persistent AGN. Using the flux measured by \citet*{Lang14b}, we find $W1-W2=0.8\pm 0.1$, similar to colors of low-redshift quasars \citep{Stern12}. Furthermore, the mid-infrared light curve of this source, derived by including the NEOWISER catalog \citep{Mainzer14}, shows variability both before and after the discovery of the optical transient, which is unlike the observed WISE light curves of other TDFs in our sample \citep{vanVelzen16b,Jiang16}. 

\subsubsection{iPTF}
Three flares in our sample originate from iPTF, which is the successor of PTF: iPTF-15af (N. Blagorodnova et al. in prep) iPTF-16axa \citep{Hung17}, and iPTF-16fnl \citep{Blagorodnova17}. The iPTF search was conducted with the same telescope and camera as PTF, but cadence and follow-up strategy were different. Contrary to the PTF search by \citet{Arcavi14}, the three flares from iPTF were not selected based on their luminosity, but based on their color and spectral similarity to previous TDFs. For iPTF we adopt $m<19.5$ in the $r$ band as the effective flux limit. For the flare iPTF-16fnl, we use the blackbody temperature reported by \citet{Brown18}.

\subsubsection{ASAS-SN}
Four flares in our sample originate from ASAS-SN \citep{Shappee14}: ASASSN-14ae \citep{Holoien14}, ASASSN-14li \citep{Holoien16}, ASASSN-15oi \citep{Holoien16b}, and ASASSN-15lh \citep{Dong16}. The nature of the fourth flare, ASASSN-15lh, is controversial: both a supernova \citep{Dong16,Godoy-Rivera17} and a TDF \citep{Leloudas16,Margutti17} have been proposed. In this paper we will consider both possible origins separately. For ASAS-SN we adopt an effective flux limit that is similar to the image flux limit, $m<17.3$ in the $g$ band.

Two flares from this survey are outside the SDSS footprint. For ASASSN-15oi we use the Pan-STARRS catalog \citep{Flewelling16} to obtain the host photometry. For ASASSN-15lh we use the host galaxy magnitudes from the best-fit population synthesis model of \citet{Leloudas16}. The measurement of the velocity dispersion of the host galaxy of ASASSN-15lh is presented in \citet{Kruhler17}.

\begin{figure}
\includegraphics[trim=4mm 2mm 0mm 5mm, clip, width=0.48\textwidth]{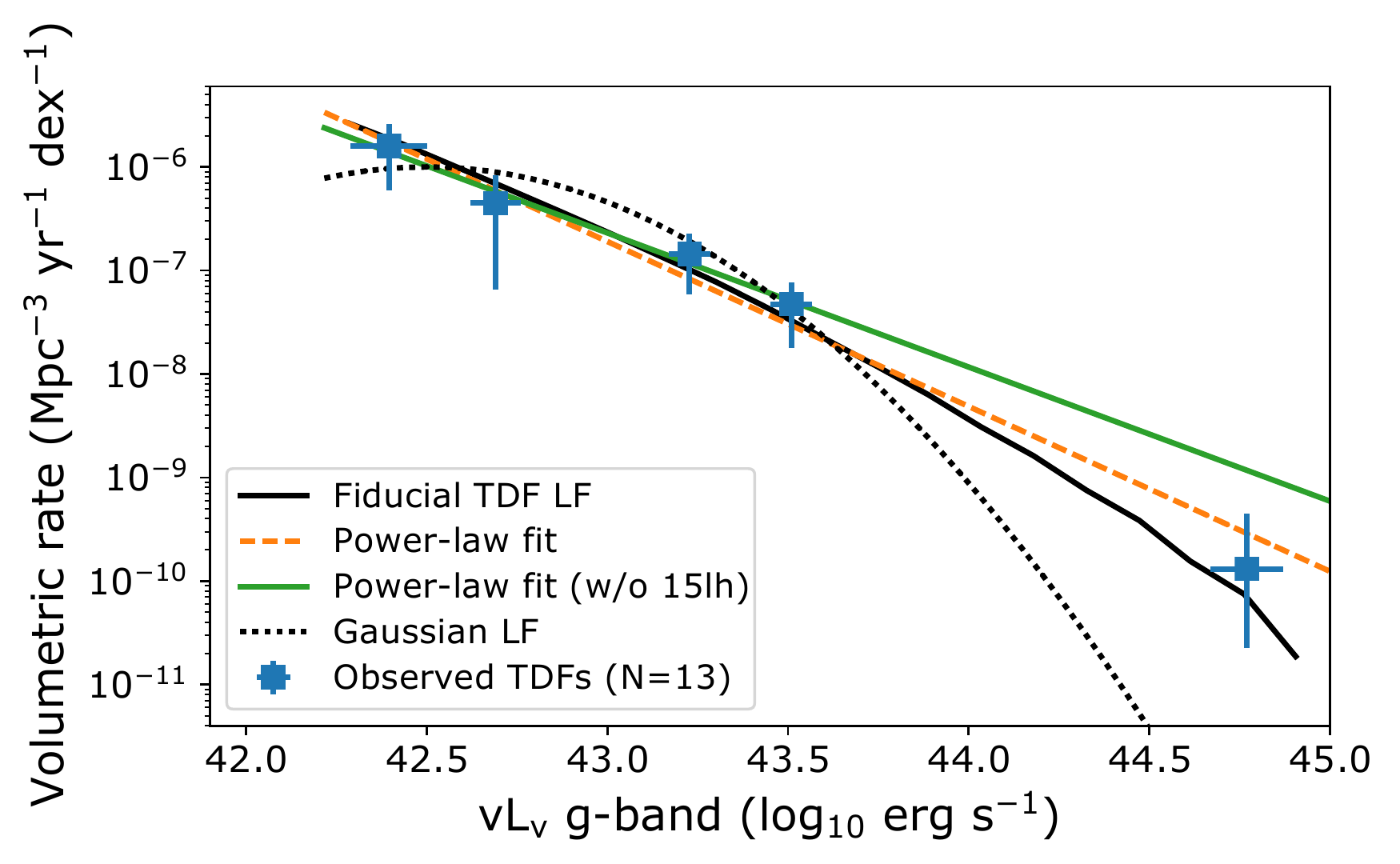} 
\caption{The TDF luminosity function (LF). The number of sources in these five bins is $\{4,2,3,3,1\}$ (low to high). The last bin contains the TDF candidate ASASSN-15lh. The dashed lines show two a power-law fits (Eq.~\ref{eq:rateVmax}) while the dotted line shows a Gaussian LF (Eq.~\ref{eq:rateVmax_gau}). The solid line presents our default model for the TDF luminosity function (see~Sec.\ref{sec:modelLFs}). }\label{fig:Lg_func}
\end{figure}

%
\section{Luminosity/Mass Functions}\label{sec:LF}
\begin{figure}
\includegraphics[trim=4mm 2mm 0mm 5mm, clip, width=0.48 \textwidth]{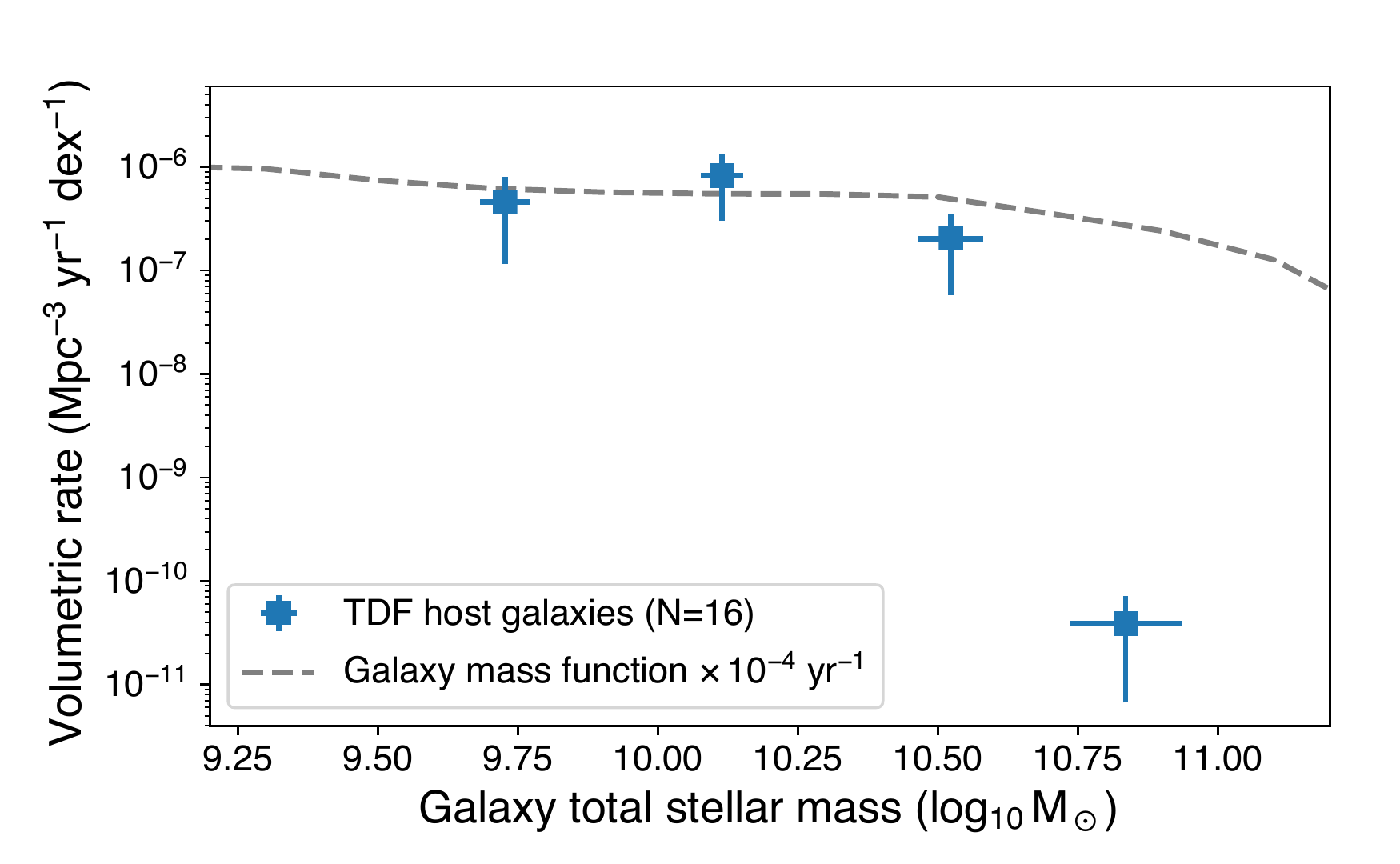} 
\caption{The TDF host galaxy stellar mass function. The number of sources in these three bins is $\{5,7,3,1\}$ (low to high). The highest-mass bin contains the TDF candidate ASASSN-15lh.  The dashed line shows a galaxy mass function \citep{Baldry12}, multiplied with a constant TDF rate of $10^{-4}$ galaxy$^{-1}$ yr$^{-1}$.}\label{fig:mass_func}
\end{figure}

%
{ For a survey of sources with a constant flux, the LF (e.g., the number of quasars per Mpc$^{3}$ as a function of luminosity) can be estimated by weighting each source by the maximum volume, $V_{\rm max}$, in which the source can be detected \citep{Schmidt68}. For transients, we are interested in the volumetric rate (e.g., the number of SNe per cubic Mpc per year as a function of peak luminosity). To estimate the volumetric rate from a survey for transients, we can also use the ``1/$V_{\rm max}$'' method, but now the weight should include the duration of the survey. We therefore define 
\begin{equation}
\mathcal{V}_{\rm max} \equiv V(z_{\rm max})~A_{\rm survey}\times\tau_{\rm survey} \quad.
\end{equation}
Here $A_{\rm survey}\times \tau_{\rm survey}$ denotes the product of the effective survey duration and survey area, and $V(z_{\rm max})$ is the volume (per unit solid angle) corresponding to the maximum redshift. As explained in Sec.~\ref{sec:sample}, the product of the effective survey duration and survey area follows from the detected number of TDF candidates, while the maximum volume follows from the survey flux limit and the peak luminosity of the transient.} In Figs.~\ref{fig:Lg_func} and \ref{fig:mass_func} we show $1/\mathcal{V}_{\rm max}$ binned by the peak $g$-band luminosity and galaxy mass, respectively. 

Since the PTF search for TDFs \citep{Arcavi14} used a luminosity selection (see~Sec.~\ref{sec:PTF}), we exclude these events when we compute the rate as a function of $L_{g}$. We also have to exclude iPTF-15af since the photometric data of this flare have not been published yet. We are thus left with $17-4=13$ sources. For TDFs with a measurement of the host galaxy velocity dispersion, the black hole mass is estimated from the $M$--$\sigma$ relation \citep{Ferrarese00,Gebhardt00}. We adopt the relation of \citet{Gultekin09}, obtained using both early-type and late-type galaxies. For the TDF host galaxies that have measured velocity dispersions, we compute the maximum volume using the lower value of $z_{\rm max}$ from the flux limit for the detection of the flare and the host galaxy flux limit for measuring the velocity dispersion---the former is the limiting factor for most sources (see the last column of Table~\ref{tab:TDFs} and Table~\ref{tab:hosts}).

The uncertainty on each bin of $\sum 1/\mathcal{V}_{\rm max}$ is estimated from $ \sum 1/\mathcal{V}_{\rm max}^{2}$ \citep{Schmidt68}. This yields a typical uncertainty of 0.3~dex for each bin, which is comparable to the Poisson uncertainty. For bins that contain only one source, we compute the uncertainty on the volumetric rate using the 1$\sigma$ confidence interval for Poisson statistics, [0.17,  3.41].  

The sum of $1/\mathcal{V}_{\rm max}$ for all 13 TDF candidates that we use for the LF yields a rate of $(8\pm 4) \times 10^{-7}\, {\rm Mpc}^{-3}{\rm yr}^{-1}$. 

The volumetric rate as a function of $L_{g}$ (Fig.~\ref{fig:Lg_func}) shows a steep decrease that can be parameterized as 
\begin{equation}\label{eq:rateVmax}
\frac{d \dot{N}}{d \log_{10}L} = \dot{N}_{0}  ~ (L / L_{0})^{a}
\end{equation}
For $L_{0}=10^{43}\,{\rm erg}\,{\rm s}^{-1}$, a least-squares fit yields $\dot{N}_{0} = (1.9 \pm 0.7) \times 10^{-7}\, {\rm Mpc}^{-3}\,{\rm yr}^{-1}$ and $a=-1.6 \pm 0.2$. If we exclude the luminous TDF candidate ASASSN-15lh, we find a more shallow slope with a larger uncertainty: $a=-1.3\pm 0.3$ and $\dot{N}_{0} = (2.3 \pm 0.8 )\times 10^{-7}\, {\rm Mpc}^{-3}\,{\rm yr}^{-1}$. When excluding ASASSN-15lh, a Gaussian function    
\begin{equation}\label{eq:rateVmax_gau}
\frac{d \dot{N}}{d \log_{10}L} = \dot{N}_{0'} \,  \exp[-(\log_{10}(L / L_{0'})^{2} /2b^{2} ] \quad
\end{equation}
with $L_{0'}=10^{42.5} \, {\rm erg}\,{\rm s}^{-1}$, $b=0.4$, and $N_{0'}=1\times 10^{-6}\, {\rm Mpc}^{-3}\,{\rm yr}^{-1}$, also provides a reasonable description of the LF (Fig.~\ref{fig:Lg_func}).

To convert our measurement of the volumetric TDF rate to a rate per galaxy, we compute the volumetric rate as a function of total stellar mass and divide by the stellar mass function of \citet{Baldry12}. For a stellar mass in the range $10^{9.5}<M_{\rm galaxy}/M_{\odot}<10^{10.5}$ a constant rate of $10^{-4}$ galaxy$^{-1}$ yr$^{-1}$ is consistent with our observations (Fig.~\ref{fig:mass_func}). 

\begin{figure}
\includegraphics[trim=4mm 2mm 0mm 5mm, clip, width=0.48 \textwidth]{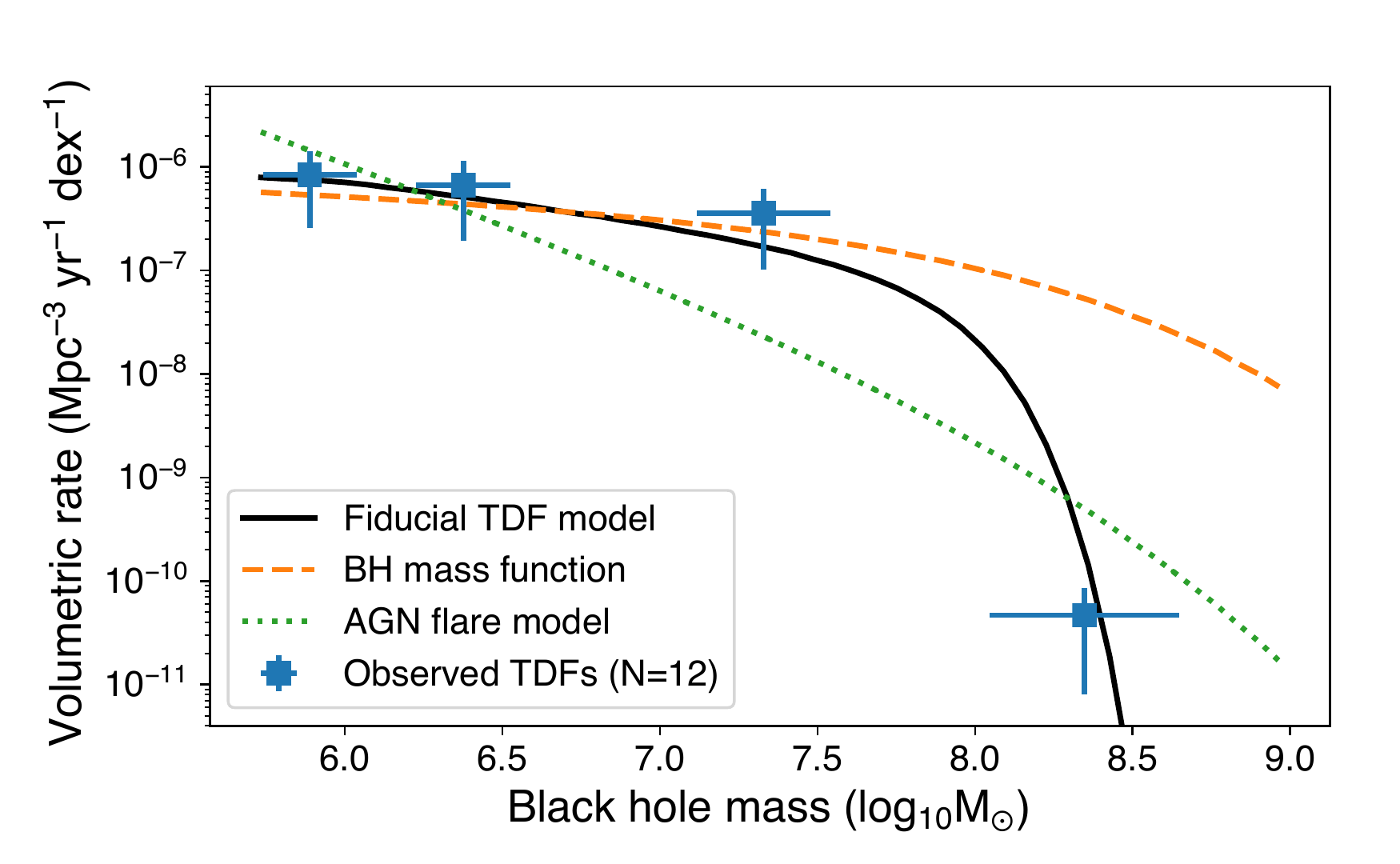} 
\includegraphics[trim=4mm 2mm 0mm 5mm, clip, width=0.48 \textwidth]{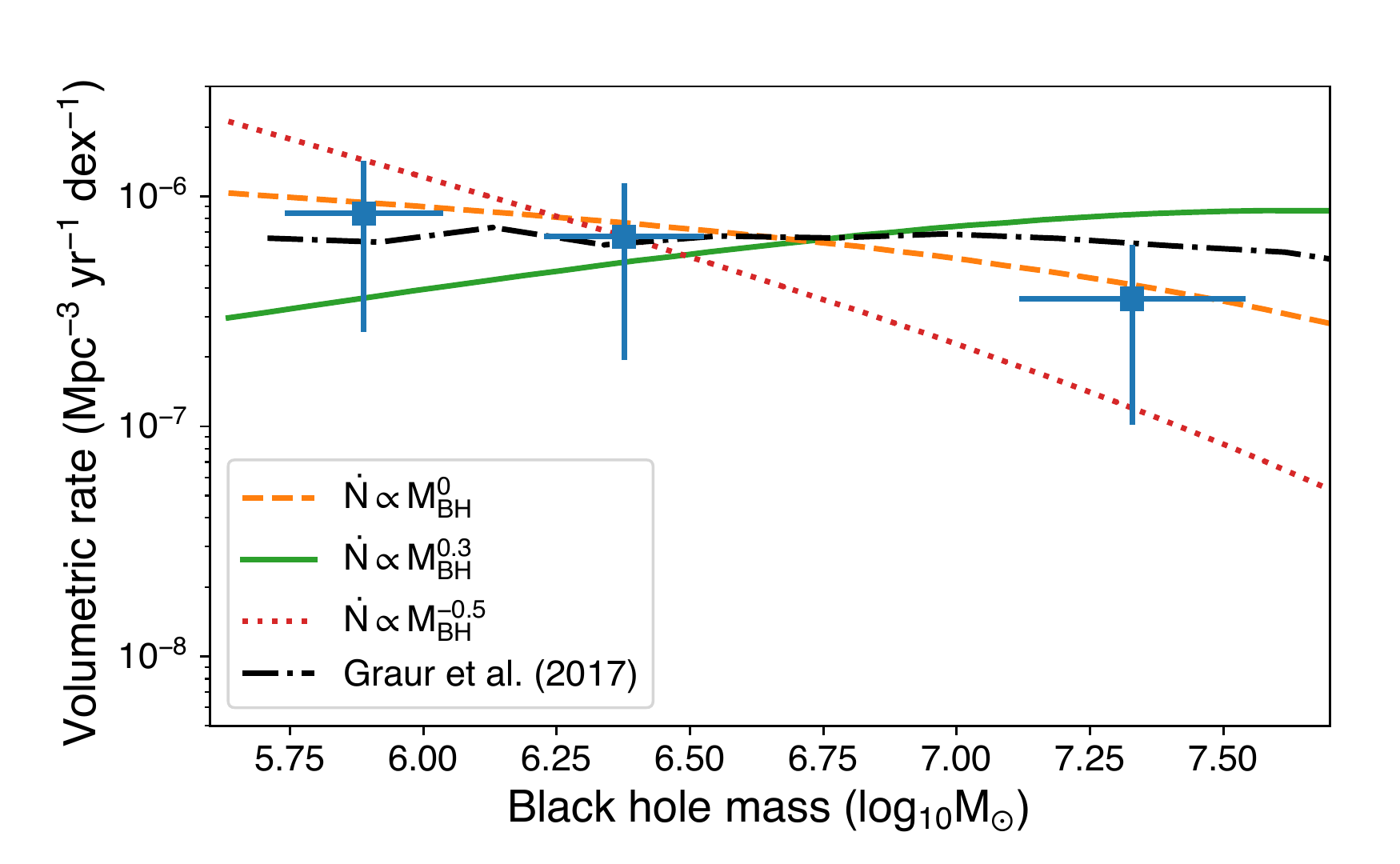} 
\caption{The TDF host galaxy black hole mass function. The number of sources in these four bins is $\{5,4,2,1\}$ (low to high). The highest-mass bin contains the TDF candidate ASASSN-15lh. In the top panel, the dashed line shows the \citet{Shankar04} black hole mass function multiplied with a constant TDF rate of $6\times 10^{-5}$ per black hole per year. The solid line shows the result of using this mass function as input to our model of the TDF rate (Eq.~\ref{eq:modelrate}). The dotted line indicates the mass function that would be obtained if the wait time between flares scales linearly with black hole mass. In the bottom panel, we compare four different predictions for the scaling of the disruption rate below the Hills mass (Sec.~\ref{sec:modeleventrates}). } \label{fig:MBH_func}
\end{figure}

The rate as a function of black hole mass (Fig.~\ref{fig:MBH_func}) is also observed to be roughly constant for $M_{\bullet}<10^{7.5}\,M_{\odot}$. 
However, the high luminosity of the TDF candidate ASASSN-15lh yields a very large $\mathcal{V}_{\rm max}$ and thus implies a rapid decrease of the volumetric rate for $M_{\bullet}\gtrsim 10^{7.5}\,M_{\odot}$. 

The decrease of the rate toward the highest-mass bin is at least 3 orders of magnitude. 
Arguably the only conceivable mechanism that can yield such an extreme turnover is the suppression of the flare rate by the black hole horizon. 
We can thus conclude that {\it if} ASASSN-15lh is a member of the TDE family, the population of observed TDFs as a whole is consistent with the predicted suppression of the rate due to the direct capture of stars by the black hole. 
However, if ASASSN-15lh is not due to a TDE, the mass function of the remaining TDFs in our sample is not a useful tool to measure rate suppression. Instead, we need to compare the observed mass distribution to the expected distribution in a flux-limited sample. This requires a forward-modeling approach, which is explained in the next section.

%
\section{Forward Modeling}\label{sec:mock}
%
In the previous section we used the $1/V_{\rm max}$ method to reconstruct the TDF LF and mass function. In this section, we start with a model for the flare luminosity function and event rate and try to reproduce the observed distribution of luminosity and host galaxy mass. This forward-modeling approach has two advantages over a $1/V_{\rm max}$ reconstruction. First of all, we can include additional selection criteria beyond the survey flux limit (e.g., the contrast between the flux of the host and flare). Second, we can assign a significance to the apparent lack of events from high-mass black holes.  

Our forward-modeling method consists of four steps: (i) draw flares with a peak luminosity from a model LF; (ii) insert these flares into a flux-limited galaxy sample; (iii) assign each flare a weight based on the event rate in its host galaxy; (iv) sum these weights for the simulated flares that pass the requirement for detection in each survey. In the following four subsections we provide the details of these steps.


\begin{figure}
\includegraphics[trim=4mm 2mm 0mm 5mm, clip, width=0.48 \textwidth]{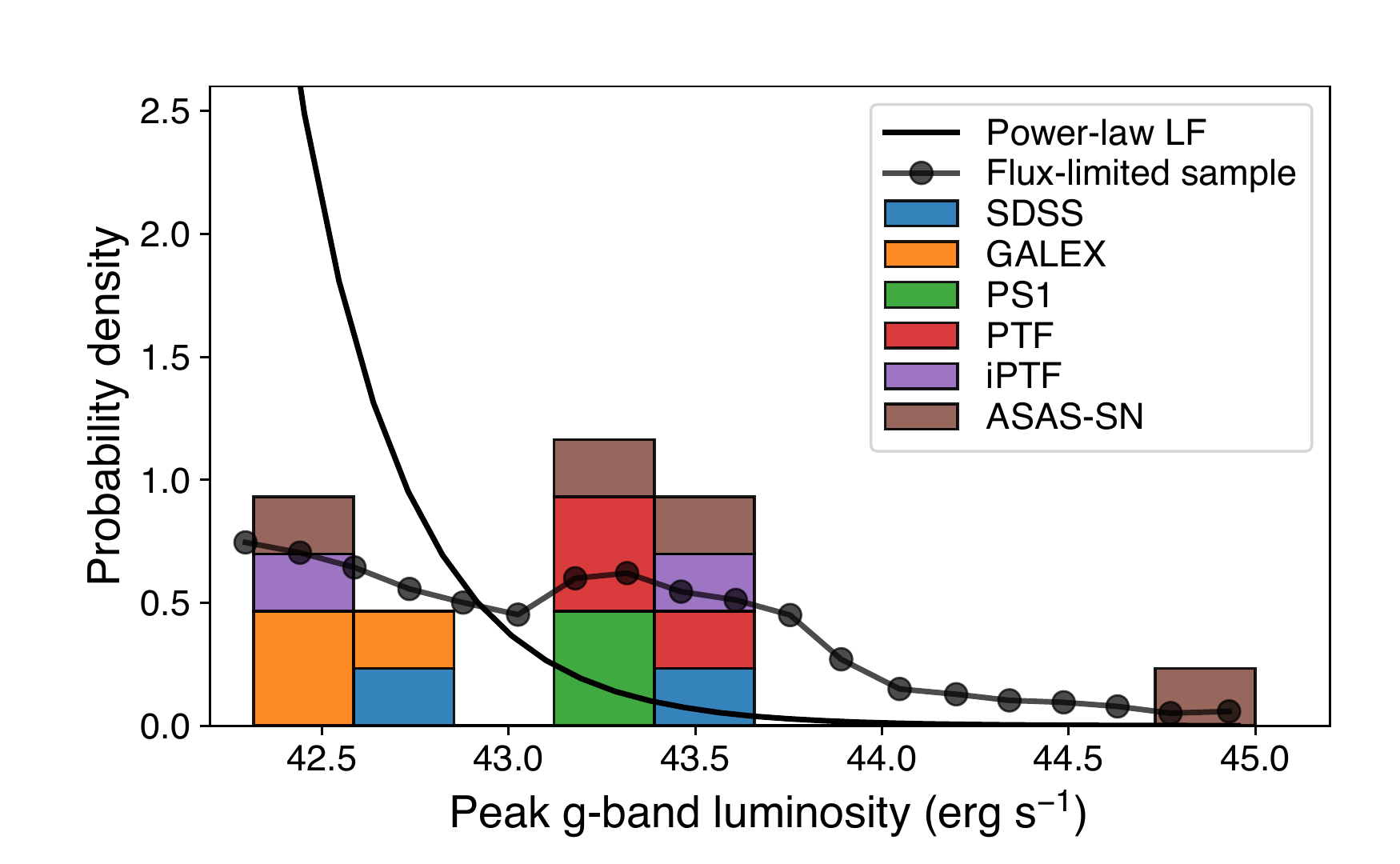}
\caption{Demonstrating the difference between a luminosity function (LF) and a flux-limited sample. The solid line shows the LF obtained from the $1/V_{\rm max}$ method (Fig.~\ref{fig:Lg_func}). The dashed line shows the result of our forward analysis: a mock TDF sample obtained after drawing flares from a power-law LF and applying the selection criteria of each survey. }\label{fig:Lg} 
\end{figure}

\subsection{Model Luminosity Functions}\label{sec:modelLFs}
Ideally, a model for TDFs or TDF impostors would yield a prediction for the LF of these events that can be tested using the observed luminosity distribution (Fig.~\ref{fig:Lg}). However, these models are not yet mature enough to predict an LF from first principles. We therefore use a more empirical approach and only consider LFs that are known to reproduce the observed luminosity distribution. We will consider two different LFs: one for SNe and one for AGN flares and TDFs. 

The observed LF from the 1/$V_{\rm max}$ method (Fig.~\ref{fig:Lg_func}) provides a good starting point for our empirical LF models. Indeed, if we draw TDFs from a power-law LF with $a=-1.5$ (see Eq.~\ref{eq:rateVmax}) and apply the survey selection criteria (see Sec.~\ref{sec:selection}) we reproduce the observed distribution of $L_{g}$, see Fig.~\ref{fig:Lg}. We therefore use this power law as the model LF of nuclear SNe.

A simple power law is unlikely to provide a correct description of the LF of transients that are due to massive black holes, i.e., TDFs or AGN flares. Due to the  abundance of low-mass galaxies, a power-law LF will yield too many transients with super-Eddington luminosities.
Motivated by the observation that the observed distribution of the Eddington ratio peaks near unity (Fig.~\ref{fig:fluxlim_fEdd}), we define the LF for AGN flares and TDFs as follows. We draw from the same power law that is used to model SNe, but we only accept flares with an Eddington ratio in the range $-3>\log_{10}(f_{\rm Edd})<0.3$. 
{Here $f_{\rm Edd}\equiv  L_{\rm bol} / L_{\rm Edd}$, with $L_{\rm Edd}=1.3\times 10^{38}\, M_{\bullet}$\,erg\,s$^{-1}$}. To assign a bolometric luminosity ($L_{\rm bol}$) to each simulated flare, we compute the blackbody luminosity by drawing a blackbody temperature from a lognormal distribution centered on $2.5\times 10^{4}$~K with a standard deviation of 0.15~dex (which provides a good description of the observed blackbody temperatures of the TDF candidates in our sample). The resulting LF for TDFs and AGN flares using this approach is shown in Fig.~\ref{fig:Lg_func}.

The next step is to insert the simulated flares into a host galaxy sample. 

\subsection{Synthetic Galaxy Sample}\label{sec:syngal}
While large galaxy surveys like SDSS provide a good census of galaxy properties at $z<0.1$, many TDFs are found at higher redshift, where galaxy properties are more difficult to observe directly. We therefore need to construct a synthetic galaxy sample.  

We use the galaxy LF measured by \citet{Cool12} to find the number of galaxies in bins of redshift and absolute magnitude, for blue and red galaxies separately. We populate each bin using the properties of real galaxies from the NYU-Valued-Added Galaxy Catalog \citep[NYU-VAGC;][]{Blanton05}, again for red and blue galaxies separately. In each bin, we compute the apparent magnitude using the median k-correction of the galaxies in that bin. We keep only bins with an apparent magnitude $m_{r}<22$. 

Because we use a redshift-dependent LF for blue and red galaxies \citep{Cool12}, our synthetic flux-limited galaxy sample contains most effects of galaxy evolution (e.g., the increase of the quiescent-galaxy density to lower redshift). Our approach also accounts for the fact that blue galaxies are easier to detect at higher redshift (due to the k-correction). And finally, because we use real galaxies to populate each absolute magnitude bin, correlations of galaxy properties (e.g., mass or size) with luminosity are part of our sample. We confirmed that the median redshift of our synthetic galaxy sample, $\left<z\right>=0.47$, is consistent with the median photometric redshift of real galaxies (also selected with $m_{r}<22$) in the co-add of SDSS Stripe 82 \citep{Reis12,Annis14}. 

The total stellar mass of each galaxy in NYU-VAGC is estimated from the 2MASS and SDSS broadband photometry using the \verb kcorrect  software \citep{blanton07}. To estimate the starformation rate (SFR), we use the specific SFR within the SDSS spectroscopic fiber as measured by the MPA-JHU group \citep{Kauffmann03b,Brinchmann04}. To estimate the bulge mass, we use the bulge-to-total ratio ($B/T$) measured in $r$ band by \citet{Lackner12}. These measurements are not available for all NYU-VAGC galaxies, so we assigned each galaxy in our synthetic sample a $B/T$ using the nearest match in the 3D vector space spanned by $M_{r}$, $g-r$, and $r-i$. 

The last quantity we wish to assign to our synthetic galaxy sample is the velocity dispersion. Unfortunately, the resolution of the SDSS spectrograph limits reliable measurements to $\sigma \gtrsim 100$~km~s$^{-1}$, which is larger than the typical velocity dispersion of TDF host galaxies \citep{Wevers17}. Following the approach of \citet{Bezanson11}, we use the virial theorem to estimate the velocity dispersion of each galaxy,
\begin{equation}\label{eq:sigma}
\sigma = \sqrt{\frac{GM}{k K(n) r_{e}}} \quad .  
\end{equation} 
Here $r_{e}$ is the effective radius, $k$ is a scale factor that accounts for the mean difference between the dynamical mass and the stellar mass estimated from the photometry \citep{Taylor10}, and $K(n)$ is a virial constant \citep{Bertin02} that depends on the Sersic index ($n$),
\begin{equation}
K(n) = \frac{73.32}{10.465 + (n-0.94)^{2}}+0.954\quad .
\end{equation} 
Using the stellar mass, effective radii, and Sersic indices reported in the NYU-VAGC \citep{Blanton05}, we find that $k=0.560$ is required to match the observed velocity dispersion to the estimate from Eq.~\ref{eq:sigma}. This calibration is consistent with the value of $k$ reported by \citet{Bezanson12}. For $\sigma>100$~km~s$^{-1}$, the scatter between the observed value of the velocity dispersion and the value from Eq.~\ref{eq:sigma} is 0.08~dex. 

The bulge mass and velocity dispersion can be used to estimate the mass of the black hole at the galaxy's center. We use the \citet{Gultekin09} $M$--$\sigma$ relation for their sample of ``all'' galaxies (i.e., both early-types and late-types):
\begin{equation}\label{eq:gulte_sigma}
\log_{10} M_{\bullet} =  8.13 + 4.24 \log_{10}(\sigma / 200~{\rm km}\,{\rm s}^{-1}) \quad.
\end{equation}
To estimate the black hole mass from the bulge mass, we adopt the \citet{Gultekin09} $M$--$L_{V}$ relation, and we use the NYU-VAGC galaxies to measure the small correction to the power-law index due to the luminosity dependence of the mass-to-light ratio, which yields 
\begin{equation}\label{eq:gulte_bulge}
\log_{10} M_{\bullet} =  8.40 + 1.16 \log_{10}(M_{\rm bulge}/10^{11} M_{\odot}) \quad.
\end{equation}
We apply Gaussian noise with a standard deviation of 0.4~dex \citep{Gultekin09} when assigning the black hole mass based on the host galaxy properties. We find that the two methods to estimate the black hole mass agree reasonably well; the difference between using the velocity dispersion and the galaxy bulge mass roughly scales as $0.2 \log_{10}(M-7.5)$~dex, with $M$ the mass from the $M$--$\sigma$ relation. Since reliable $B/T$ measurements are not available for most of the TDF candidates in our sample, we will use the black hole mass estimate from the $M$-$\sigma$ relation as the default value in our analysis.

We simulated $10^{7}$ galaxies. This sample is available online (see Table~\ref{tab:syngal}).

\begin{deluxetable}{l || c  c}
\tablecolumns{3}
\tablewidth{0pt}
\tablecaption{
modeling scenarios} 
\startdata
	& $\dot{N}\propto L^{-2.5}$   &  $-3 <\log_{10} f_{\rm Edd} < 0.3$  \\[3pt]
	\hline\hline \\[2pt]
	$\dot{N}\propto {\rm const}$     				& SNe 	& AGN flares \\[2pt]
	$\dot{N} \propto {\rm SFR} $ 			& SNe 	& -- \\[2pt] 
	$\dot{N} \propto {\rm mass} $ 			& SNe 	& -- \\[2pt] 
	$\dot{N} \propto M_{\bullet}^{-1}$ 	& -- 		& AGN  flares \\[2pt]
	Eq.~\ref{eq:modelrate} 								& --  	 	& TDFs	
	\enddata
\tablecomments{The cells of this matrix show which combinations of event rate (rows) and luminosity function (columns) are considered as a description for AGN flares, nuclear SNe, or TDFs. The two luminosity functions have the same power-law index, but the function used for TDFs and AGN flares is capped based on  the ratio of the blackbody luminosity to the Eddington luminosity ($f_{\rm Edd}$). } \label{tab:modelmatrix}
\end{deluxetable}

\begin{figure*}
\includegraphics[trim=4mm 2mm 0mm 5mm, clip, width=0.48 \textwidth]{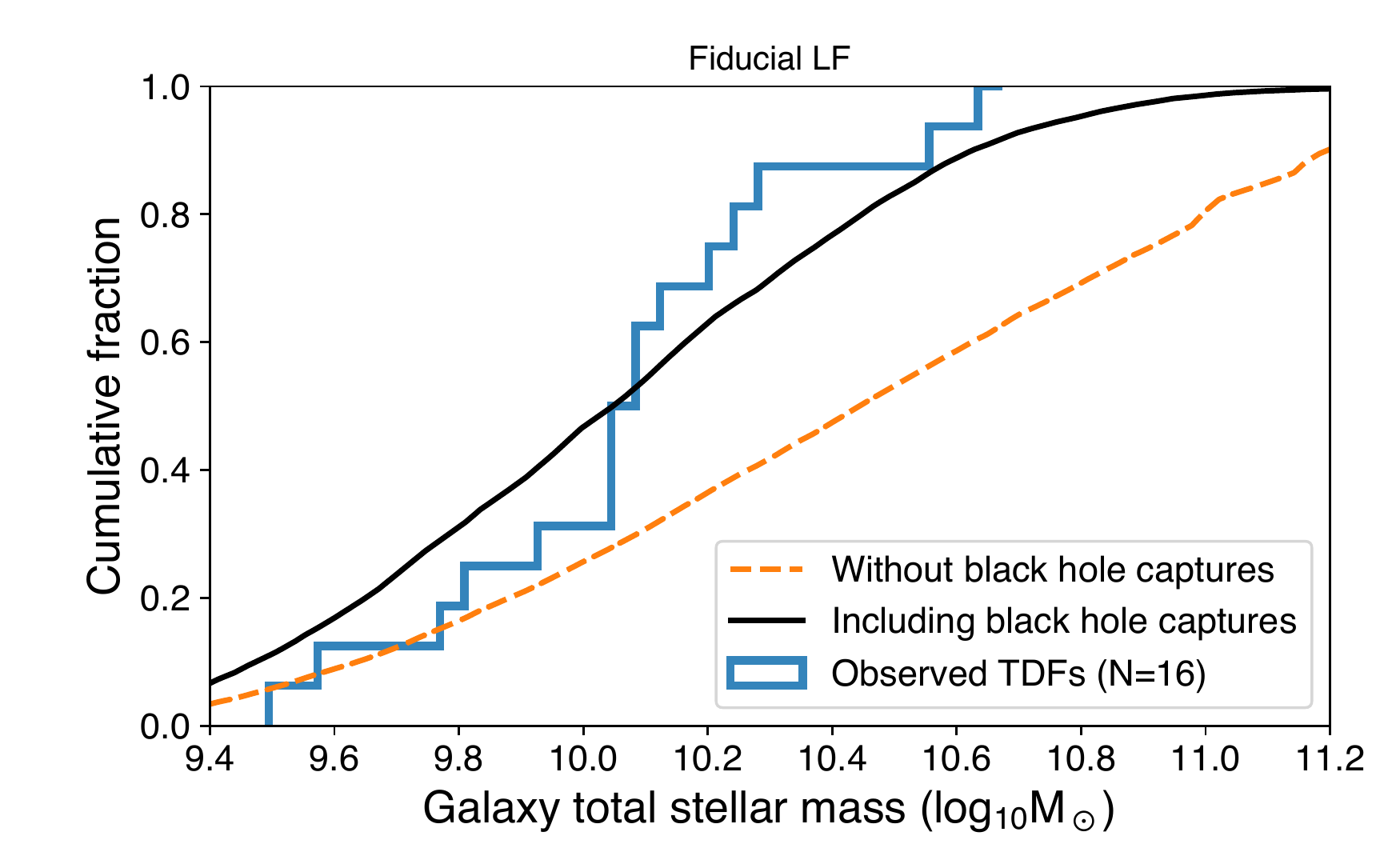} 
\includegraphics[trim=4mm 2mm 0mm 5mm, clip, width=0.48 \textwidth]{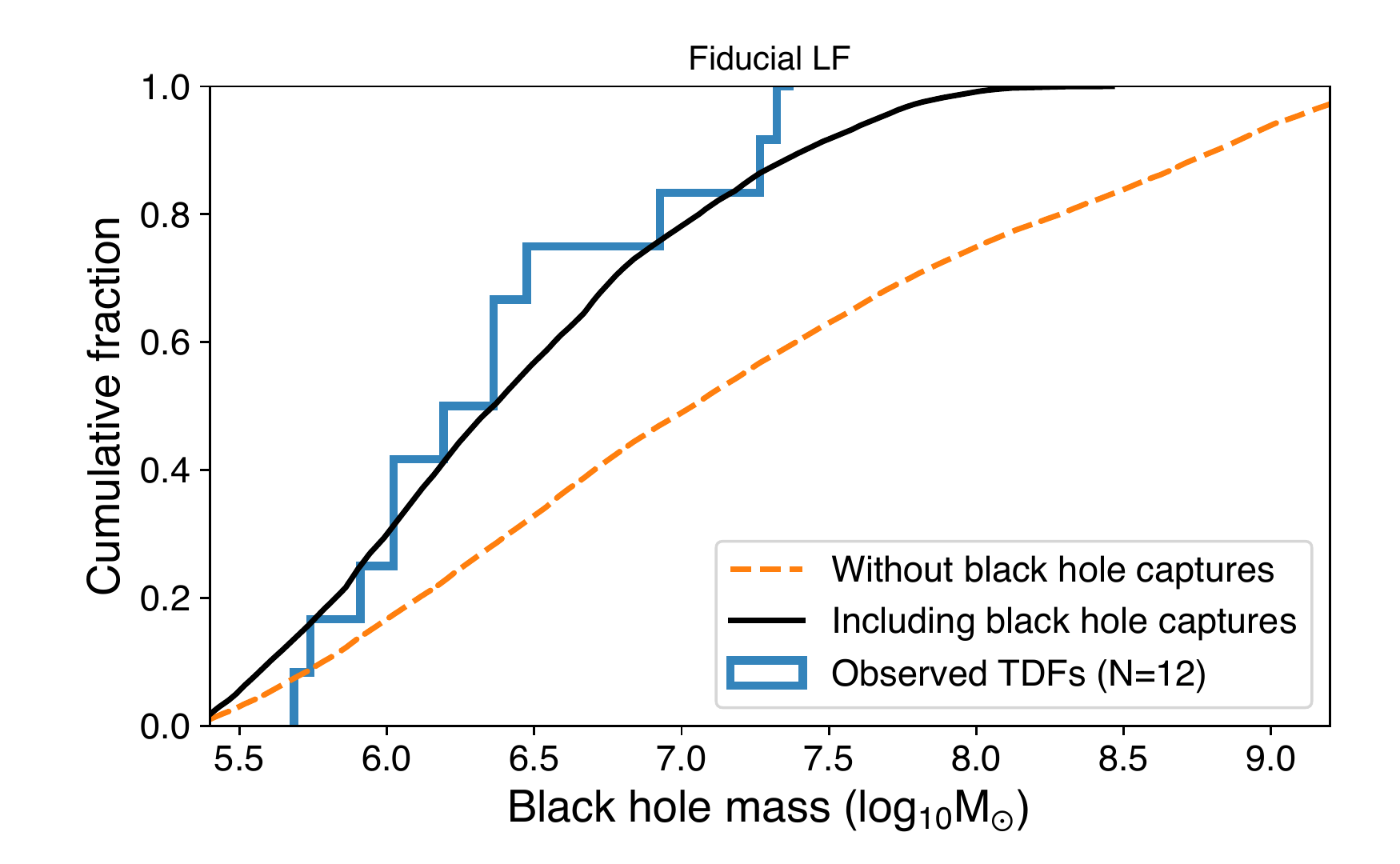} 
\caption{Cumulative distribution of host galaxy stellar mass and black hole mass. We show the observed distribution, compared to the distribution for two different mock TDF samples. The black solid line is our fiducial TDF model, using Eq.~\ref{eq:modelrate} to account for the suppression of the TDF rate due to the capture of stars by black holes. The dashed line shows the distribution that is obtained if the event rate is independent of mass.  This second scenario clearly is inconsistent with the observations, as it predicts too many flares from high-mass host galaxies. }\label{fig:fluxlim}
\end{figure*}

\begin{figure*}
\includegraphics[trim=4mm 2mm 0mm 5mm, clip, width=0.48 \textwidth]{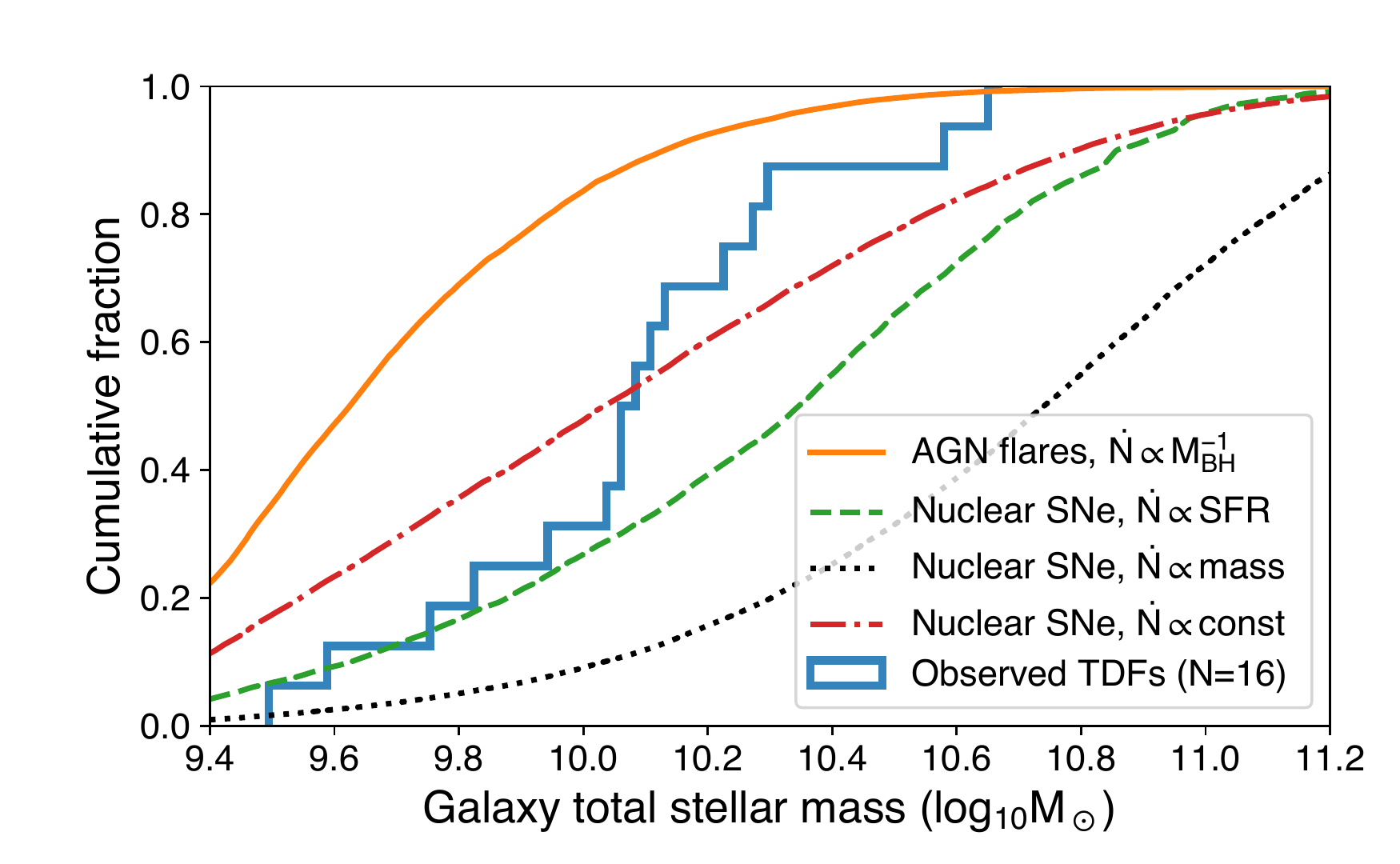}
\includegraphics[trim=4mm 2mm 0mm 5mm, clip, width=0.48 \textwidth]{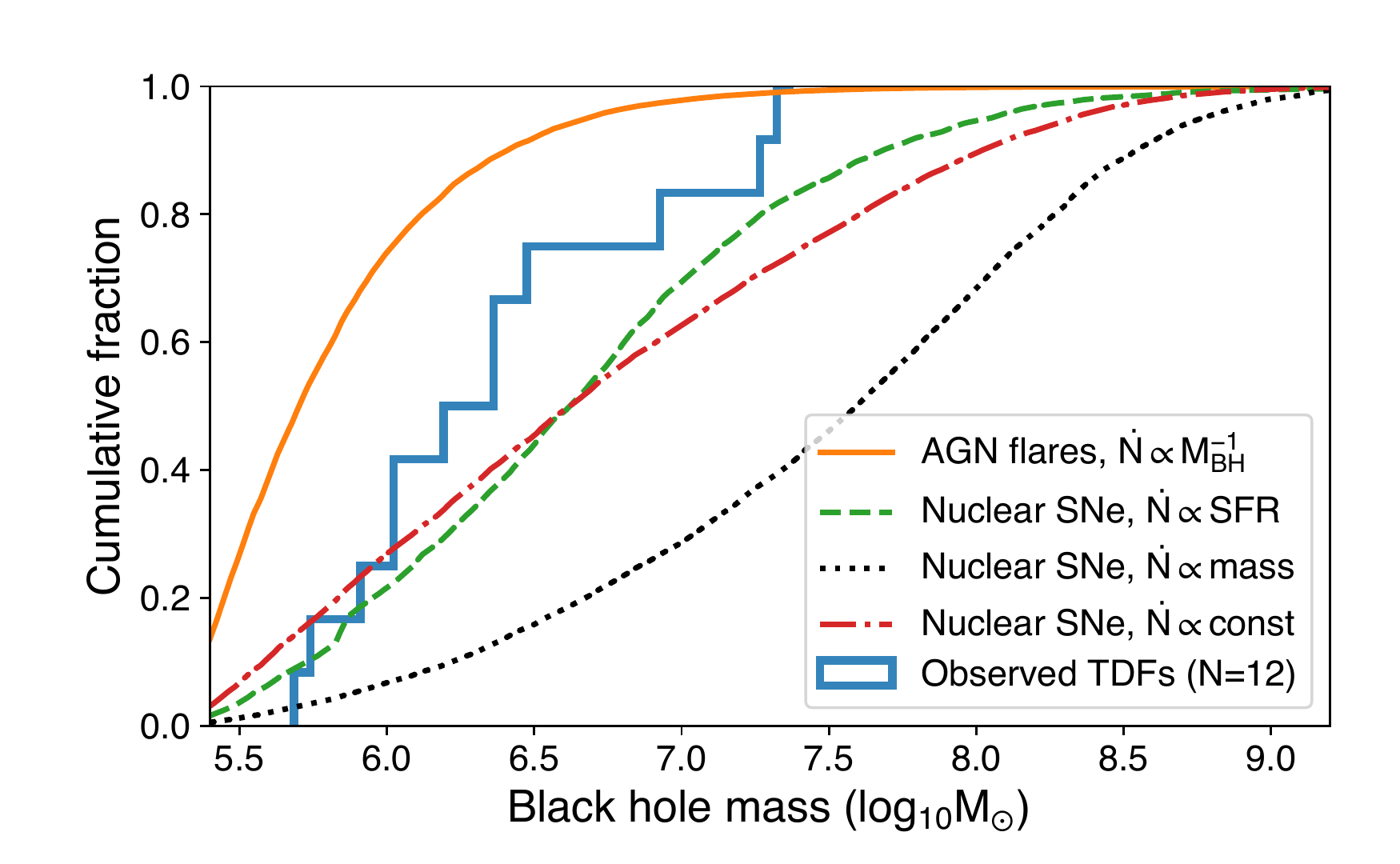}
\caption{Identical to Fig.~\ref{fig:fluxlim}, but showing three more models for the event rate and flare luminosity function (see Table~\ref{tab:modelmatrix}). We see that both the AGN flare scenario and SNe that trace the starformation rate (SFR) or galaxy mass are inconsistent with the observed distributions. While an SN model with a rate that is independent of galaxy properties is consistent with the observed mass distribution, this scenario is not consistent with the observed distribution of the Eddington ratio (Fig.~\ref{fig:fluxlim_fEdd}). }\label{fig:fluxlim_other}
\end{figure*}

\subsection{Model Event Rates}\label{sec:modeleventrates}
We consider four possible models for the scaling of the event rate with galaxy properties. First of all, the simplest assumption is a galaxy-independent rate. Next, we consider an event rate proportional to the SFR. This scaling could be expected if current optical TDF candidates are due to a new type of stellar explosion in galactic nuclei \citep{Saxton16}. 

To model flares caused by AGN disk instabilities, we consider an event rate that is inversely proportional to central black hole mass, $\dot{N} \propto M_{\bullet}^{-1}$ (i.e., the wait time between outbursts is proportional to the black hole mass; see Sec.~\ref{sec:insta}). Finally, to estimate the rate of flares due to TDEs, we use the following equation:
\begin{equation}\label{eq:modelrate}
	\dot{N} \propto M_{\bullet}^{\beta} e^{- (M_{\bullet}/10^{8}M_{\odot})^{2} } \quad. 
\end{equation}
The power-law index $\beta$ parameterizes how the disruption rate changes due to the dynamics of the host galaxy; predictions for this index range from $+0.3$ \citep{Brockamp11}, to $-0.2$ \citep{Wang04,Kochanek16}, and $-0.5$ \citep{Stone16b}. All of these predictions are broadly consistent with the observed mass function derived from the $1/V_{\rm max}$ method, but steeper power laws can be ruled out (see Fig.~\ref{fig:MBH_func}, bottom panel). For our fiducial TDF model, we adopt $\beta=-0.2$, the relation predicted for an isothermal sphere \citep[cf.][Eq. 29]{Wang04}. Our parameterization of the turnover in the TDF rate approximates the curve of \citet[][, Figure~4]{Kesden12b} for a solar-type star disrupted by a black hole with a spin of $a=0.9$. 
{ Finally, in Fig.~\ref{fig:MBH_func} we also compare our estimate of the TDF rate as a function of mass with the result of \citet{Graur17}, who conclude that this rate is proportional to $\Sigma/\sigma$, with $\Sigma$ the galaxy surface mass density. For this comparison, we computed $\Sigma/\sigma$ for the galaxies in our mock sample and binned the result as a function of black hole mass. }

\subsection{Mock TDF Samples}\label{sec:selection}
The last step of our forward-modeling analysis is applying the selection criteria of the surveys. Summing the event rate of simulated flares that pass the selection criteria yields the final output: a mock version of our flux-limited TDF sample (Table~\ref{tab:TDFs}). 

Besides the obvious requirement that the peak flux of the simulated flare is larger than the effective survey flux limit, we also require that the magnitude difference between the simulated flare and the host galaxy ($m_{\rm peak} - m_{\rm host}$) is less than 3~mag (which corresponds to the lowest host-flare contrast in our sample of candidate TDFs). For simulated flares observed by the PTF survey we also apply their luminosity cut ($-19>M_{r}>-21$;see Sec.~\ref{sec:PTF}). 
Similar to our method for normalizing the effective area in the $1/V_{\rm max}$ analysis, we require that the ratio of the simulated flares detected by each survey is equal to the ratio of detected TDF candidates for these surveys (see Table~\ref{tab:surveys}). 

We now compare the observed distribution\footnote{In Figs. \ref{fig:fluxlim}--\ref{fig:fluxlim_fEdd} we show the observed distributions without ASASSN-15lh.} 
of galaxy mass or black hole mass of the host galaxies of our TDF candidates with the distribution obtained for a simulated sample with and without a correction for captures (Fig.~\ref{fig:fluxlim}). We find that the simulation without a correction for captures overpredicts the number of flares from high-mass black holes. Using a Kolmogorov--Smirnov (KS) test, the hypothesis that the host galaxy stellar mass of the mock sample without captures and the host galaxy mass of observed candidate TDFs are drawn from the same distribution can be rejected with $p=2\times 10^{-3}$. If we apply the same test to the distribution of black hole mass, we again reject the null hypothesis, but with slightly lower significance ($p=2\times 10^{-2}$, due to the smaller sample size). Repeating this exercise using a Gaussian LF (Eq.~\ref{eq:rateVmax_gau}) instead of an power-law LF (Eq.~\ref{eq:rateVmax}) does not change the significance of the detection of rate suppression by black hole event horizons. 

Since we use the number of detected TDF candidates in each survey to normalize the contribution of the different surveys to the final sample of mock TDFs, the small number of flares in each survey introduces a statistical uncertainty that is not captured in a single KS test. To estimate this uncertainty, we compute multiple mock samples, drawing the number of candidate TDFs in each surveys from a Poisson distribution centered on $N_{\rm TDF}$. For each of these samples, we compute the $p$-value for rejecting the null hypothesis. The distribution of the resulting $p$-values follows a lognormal distribution with a standard deviation of only 0.2~dex. We can thus conclude that Poisson fluctuation in the number of detected TDF candidates will not lead to a false detection of horizon suppression (a 9$\sigma$ fluctuation of $N_{\rm TDF}$ is required to reach $p>0.1$).

The simulated distributions of galaxy mass and black hole mass for the other three scenarios that we consider (see Table~\ref{tab:modelmatrix}) are shown in Fig.~\ref{fig:fluxlim_other}.

%
\section{Discussion}\label{sec:discussion}
%
\subsection{TDFs Are Not Due to Stellar Explosions}
\begin{figure}
\centering
\includegraphics[trim=4mm 2mm 0mm 5mm, clip, width=0.48 \textwidth]{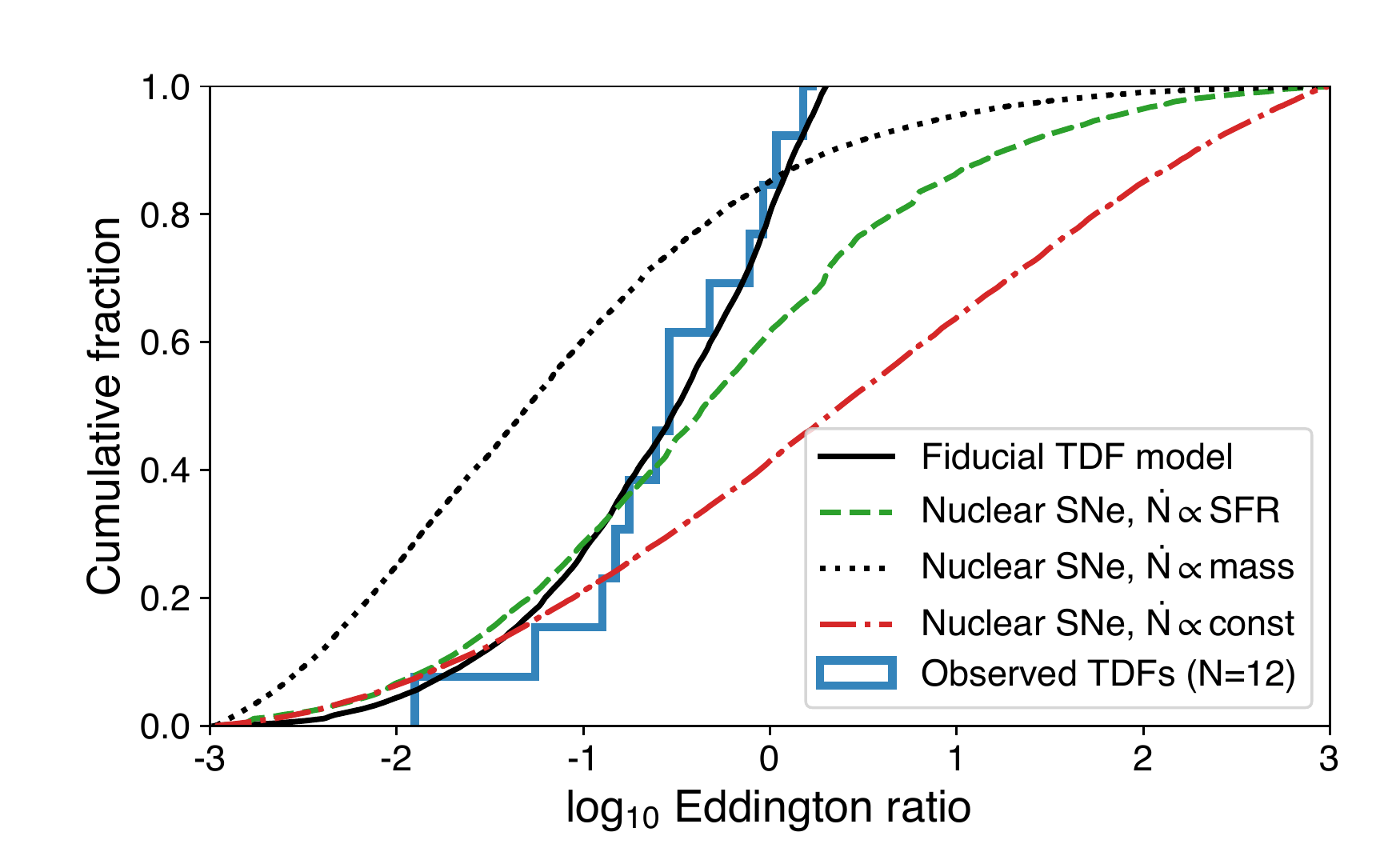}
\caption{Cumulative distribution of the Eddington ratio.  We show the observed distribution for TDF candidates with black hole mass measurements based on the host galaxy velocity dispersion, compared to the distribution predicted for a TDE scenario and three different SNe scenarios (see Table.~\ref{tab:modelmatrix}). If observed TDF are due to SNe, we would not obtain a luminosity distribution that is capped near the Eddington luminosity. }\label{fig:fluxlim_fEdd}
\end{figure}

We find that our fiducial TDF model correctly reproduces the distribution of host galaxy total stellar mass, host galaxy black holes mass, and Eddington ratio (see Fig.~\ref{fig:fluxlim} and Fig.~\ref{fig:fluxlim_fEdd}). 

From Fig.~\ref{fig:fluxlim_other}, we conclude that if the observed TDFs are due to a hypothetical new class of nuclear SNe, the rate of these events needs to be independent of host galaxy mass or SFR. This requirement could be considered unlikely, because the rate of most types of known SNe either scales with the host galaxy surface brightness or is limited to a particular subset of galaxies \citep[e.g.,][]{Fruchter06}. 

{ The strongest} evidence against the possibility that observed optical TDF candidates are a new class of SNe is the observed distribution of the Eddington ratio. The Eddington limit for photons does not apply to stellar explosions, but for each simulated SN we can still compute the Eddington ratio based on the central black hole mass of its host galaxy. If the flare rate is independent of galaxy properties and not constrained by the Eddington limit,  more than half of the observed optical TDF candidates should have super-Eddington luminosities (Fig.~\ref{fig:fluxlim_fEdd}). The luminosity of candidate TDFs, however, is observed to be capped near the Eddington luminosity. The probability that a flux-limited SN sample would produce this skewed distribution of $f_{\rm Edd}$ is  small (KS test yields $p=6\times 10^{-4}$). We find the same result if we use a Gaussian distribution (Eq.~\ref{eq:rateVmax_gau}) to draw the luminosity of the simulated SNe. 


\subsection{TDFs Are Unlikely Due to AGN}\label{sec:insta}
An instability in an AGN accretion disk could lead to a rapid increase of the accretion rate and may therefore mimic a TDF. 
This scenario has several problems, such as the observed evolution of the broad emission lines of known candidate TDFs, which get more narrow with time, while AGN show the opposite behavior \citep{Ruan16,Holoien16}. But such problems are not insurmountable because the parameter space of AGN disk instabilities has not been fully explored yet. 
Our work provides a test of the AGN flare scenario with minimal requirements. The only model prediction that is needed is the scaling of the flare rate with black hole mass. 

The wait time between AGN outbursts depends on the black hole mass and the accretion rate. If the accretion rate normalized to the Eddington limit is constant over the mass range relevant for our TDF sample, the wait time between outbursts from active black holes is predicted to scale as $\tau \propto M_{\bullet}^{p}$, with $p \sim 1$ \citep{Mineshige90,Siemiginowsk97}. The increased wait time and longer flare duration reduce the rate of detected AGN flares from massive black holes, potentially explaining the lack of TDF candidates at high mass. However, we find that a scenario with $p=1$ predicts too many flares at the low-mass end (Figs.~\ref{fig:MBH_func} \& \ref{fig:fluxlim_other}). If instead the rate of AGN flares is independent of mass ($p=0$), a flux-limited sample of AGN flares should contain many more events from black holes with a mass $>10^{7.5}\, M_{\odot}$ (Fig.~\ref{fig:fluxlim}). We can thus conclude that most AGN flare scenarios are inconsistent with the observed mass distribution of TDF host galaxies. One caveat is that most AGN outburst models assume that the disk remains radiatively efficient between outbursts. It would be interesting to compare predictions for the rate of sub-Eddington AGN (e.g., LINERs) going into outburst. 

Besides instabilities, stellar collisions near the tidal radius could also produce TDF impostors \citep{Metzger17}. These collisions happen between stars that   accrete onto the central black hole via Roche lobe overflow, and therefore they are only possible when the Roche radius lies outside the innermost stable circular orbit. The rate of these collisions will therefore diminish above a mass scale similar to the Hills mass of TDEs and could thus explain the observed turnover in the TDF mass function (Fig.~\ref{fig:MBH_func}). However, multiple grazing collisions of the same stars are required to get a sufficiently high rate of these events, and around larger supermassive black holes, stars are often destroyed by their relativistic collision velocities. As a result, rate suppression in the stellar collision model likely occurs at an order of magnitude lower black hole mass ($M_{\bullet}\sim 10^{7},M_{\odot}$; \citealt{Metzger17}), which is inconsistent with our observations.  


\subsection{Detection of Horizon Suppression}
The only scenario that can explain both the distribution of the Eddington ratio and the distribution of black hole mass requires a roughly constant rate up to a black hole mass of $M_{\bullet} \sim 10^{7.5} \, M_{\odot}$, followed by a rapid decrease toward higher mass. This, indeed, is a fundamental prediction of the TDE paradigm. 

The location of the turnover scales with black hole spin and the density of the disrupted star. Assuming that for black holes with a mass of $\sim 10^{8} \, M_{\odot}$ most of the disrupted stars are similar to the Sun \citep[e.g.,][]{Kochanek16}, our mass function appears to imply a relatively high mean spin of black holes in this mass regime---perhaps similar to the spins inferred using observations of the iron fluorescence line in nearby AGN \citep{Risaliti13,Reynolds14}. The detection of more events similar to ASASSN-15lh will be key to making a more robust inference of black hole spin from the TDF black hole mass function.  


Our measurement of the turnover in the black hole mass function relies on the $M$-$\sigma$ relation to estimate this mass, and therefore it is subject to the systematic uncertainty associated with the calibration of this relation. However, the turnover is also clearly detected in the distribution of total stellar mass (Figs.~\ref{fig:mass_func} and \ref{fig:fluxlim}): for the event rate to be independent of galaxy mass, 50\% of the TDFs in our compilation should have a host galaxy mass $M_{\rm galaxy}>10^{10.5}\,M_{\odot}$. Instead, only 3 out of 17 (including ASASSN-15lh) are found above this host galaxy mass limit. The stellar mass of TDF host galaxies and the synthetic host galaxy sample are calculated using the same method; hence the systematic uncertainty of this result is negligible. 

At the low end of the host galaxy mass spectrum, we find no significant decrease of the number of flares compared to what is expected in a flux-limited survey. This could be considered surprising, since circularization and accretion of the stellar debris for disruptions around low-mass black holes are predicted to be less efficient \citep{Guillochon15}. Our observations thus support a constant black hole occupation fraction for $M_{\bullet}\gtrsim10^{5.5}\,M_{\odot}$. 

Post-starburst galaxies (characterized as quiescent galaxies with strong Balmer absorption lines; \citealt{Dressler83,Zabludoff96}) are overrepresented among TDF hosts \citep{Arcavi14,French16,French17,Law-Smith17,Graur17}, which could be explained by a short relaxation time caused by a high central density in these galaxies \citep{StonevanVelzen16,Graur17}. An overrepresentation of post-starburst galaxies is unlikely to significantly influence the distribution of total stellar mass of the mock TDF sample because the relative mass increase in the recent star-formation episode of these galaxies is modest, 10--50\%, (\citealt{Kaviraj07}; D. French et al., in prep). Since we use the total stellar light to estimate the galaxy mass (including the old stars as measured by the near-IR flux), a post-starburst phase of TDF host galaxies will not lead to a significant change of the total stellar mass in the mock sample. 

{ Our results are consistent with the recent work by \citet*{Lu17}, who used the lack of TDF candidates from high-mass black holes to rule out the hypothesis that supermassive black holes have a surface (at some small fraction above the Schwarzschild radius).}
 
\subsection{The Luminosity Function of TDFs} 
Our work is the first to measure the shape of the LF of optical/UV-selected TDFs. We find that a steep power law provides a good description, $dN/dL \propto L^{-2.5}$ (Eq.~\ref{eq:rateVmax}). 

About one-third of the candidate TDFs in our sample were discovered after the peak in the light curve and this introduces a systematic uncertainty to the LF. If the true peak luminosity of all of the sources discovered after maximum light is a factor of 2 higher than the observed maximum luminosity, the power-law index of the LF decreases by about 10\%.

{ While the TDF LF is steep, both the observed rate as a function of black hole mass (Fig.~\ref{fig:MBH_func}) and the peak luminosity of the flare \citep{Hung17} appear to be independent of black hole mass (for $M_{\bullet}<10^{7.5}\,M_{\odot}$). Here we speculated that this} might imply that the wide range in TDF peak luminosity is determined by the mass of the star that got disrupted. For a main-sequence mass--radius relation, the peak of the fallback rate of stellar debris is expected to scale as $M_{\rm star}^{0.8}$ \citep[e.g.,][]{Guillochon13}. If the peak luminosity is proportional to this fallback rate, a \citet{Kroupa93} initial mass function ($dN_{\rm star}/dM_{\rm star} \propto M_{\rm star}^{-2.3}$ for $M_{\rm star}\lesssim 1\, M_{\odot}$) would yield a steep decrease of the event rate with peak luminosity. Since the Hills mass is proportional to the mass of the disrupted star, this scenario implies that low-luminosity TDFs, such as iPTF-16fnl, only occur in relatively low-mass host galaxies, while high-luminosity TDFs can occur across a wider host galaxy mass range. 


The steep flare LF also explains why our measurement of the average per-galaxy rate ($\approx 10^{-4}\,{\rm galaxy}^{-1}\,{\rm yr}^{-1}$; Fig.~\ref{fig:mass_func}) is higher than the rate based on SDSS  data \citep{vanVelzen14} or ASAS-SN data \citep{Holoien16}. In the SDSS analysis, the per-galaxy rate follows from $N_{\rm TDF}/ \left<\epsilon_{\rm gal}\right>$, with $\epsilon_{\rm gal}$ the search efficiency times the number of galaxies that were surveyed. Because the average TDF rate was computed using the mean of this efficiency, the brightest flare contributes more to the per-galaxy rate (the difference in $\epsilon_{\rm gal}$ between SDSS-TDE1 and SDSS-TDE2 is factor of 4; see \citealt{vanVelzen14}, Table~1). The efficiency-weighted mean luminosity of the two SDSS flares is $\approx 10^{43.4} \,{\rm erg}\,{\rm s}^{-1}$. The rate from our LF (Fig.~\ref{fig:Lg_func}) at this relatively high luminosity is $\approx 10^{-7.3}\,{\rm Mpc}^{-3}\,{\rm yr}^{-1}$, which is consistent with the volumetric rate reported from the SDSS search. 

Our measurement of the per-galaxy flare rate (Fig.~\ref{fig:mass_func}) is consistent with the theoretical predictions of the disruption rate by \citet{Stone16b}, who find $ \sim {\rm few} \times 10^{-4}\,$galaxy$^{-1}$~yr$^{-1}$. As forecasted by \citet{Kochanek16}, it appears that the tension between earlier measurements of the TDF rate and the theoretically expected rate could simply be due to the relatively high luminosity of the TDFs in these surveys. When accounting for our observation that faint TDFs occur more frequently, the discrepancy between the observed and predicted rate disappears. 

The total rate of TDFs depends on the low-luminosity turnover of the LF, which is not constrained by our current sample of flares. However, if the bolometric luminosity of TDFs is capped near the Eddington limit, the LF should start to flatten just below the luminosity of the faintest flares in our sample (for a typical  bolometric correction of $\sim 10$, a $g$-band luminosity of $L_{g}\sim10^{42.5}\,{\rm erg}\,{\rm s}^{-1}$ exceeds the Eddington limit for black holes with $M_{\bullet}<10^{5.5}\,M_{\odot}$). For an Eddington-limited emission mechanism, the peak of the luminosity distribution in a flux-limited sample shifts to higher black hole mass \citep{Kochanek16}. This could explain why the turnover of the LF has not been detected yet. 

\section{Conclusions}\label{sec:conclusion}
Our main conclusions are as follows:

\begin{itemize}
		\item We measured the luminosity function of TDFs (Fig.~\ref{fig:Lg_func}), finding a steep decrease of the event rate with luminosity (Eq.~\ref{eq:rateVmax}).
		\item For galaxies with a stellar mass of $\sim 10^{10}\, M_{\odot}$, the observed per-galaxy rate is $\approx 1 \times 10^{-4}$~yr$^{-1}$ (Fig.~\ref{fig:mass_func}).
		\item We measured the black hole mass function of TDF host galaxies (Fig.~\ref{fig:MBH_func}), finding an approximately constant volumetric rate for $M_{\bullet}<10^{7.5}\,M_{\odot}$.
		\item The sharp decrease of the volumetric rate above $M_{\bullet}=10^{7.5}\, M_{\odot}$, as implied by the high-luminosity TDF candidate ASASSN-15lh, is consistent with the suppression of the TDF rate due to the capture of stars before they are disrupted.
		\item Rate suppression due to black hole event horizons can also be detected while remaining agnostic about the origin of ASASSN-15lh. Using forward modeling to reproduce our flux-limited TDF sample, we conclude that rate suppression at high black hole mass plus an Eddington-limited emission mechanism are both required to explained the observed distribution of galaxy mass and Eddington ratio (Figs.~\ref{fig:fluxlim}--\ref{fig:fluxlim_fEdd}). 
\end{itemize}

\acknowledgments
{\small I would like to thank Thomas Wevers, Sterl Phinney, Nick Stone, Or Graur, James Guillochon, Peter Jonker, Iris Groen, and Richard I. Anderson for useful discussions and the anonymous referee for the thoughtful comments. I am also grateful to the International Space Science Institute (ISSI) in Bern for their hospitality. This works is made possible by support of NASA through a Hubble Fellowship, HST-HF2-51350. 

This research made use of Astropy, a community-developed core Python package for Astronomy \citep{Astropy-Coll13}.

Funding for the SDSS and SDSS-II has been provided by the Alfred P. Sloan Foundation, the Participating Institutions, the National Science Foundation, the U.S. Department of Energy, the National Aeronautics and Space Administration, the Japanese Monbukagakusho, the Max Planck Society, and the Higher Education Funding Council for England. The SDSS is managed by the Astrophysical Research Consortium for the Participating Institutions: the American Museum of Natural History, Astrophysical Institute Potsdam, University of Basel, University of Cambridge, Case Western Reserve University, University of Chicago, Drexel University, Fermilab, the Institute for Advanced Study, the Japan Participation Group, Johns Hopkins University, the Joint Institute for Nuclear Astrophysics, the Kavli Institute for Particle Astrophysics and Cosmology, the Korean Scientist Group, the Chinese Academy of Sciences (LAMOST), Los Alamos National Laboratory, the Max-Planck-Institute for Astronomy (MPIA), the Max-Planck-Institute for Astrophysics (MPA), New Mexico State University, Ohio State University, University of Pittsburgh, University of Portsmouth, Princeton University, the United States Naval Observatory, and the University of Washington.

The Pan-STARRS1 Surveys (PS1) and the PS1 public science archive have been made possible through contributions by the Institute for Astronomy, the University of Hawaii, the Pan-STARRS Project Office, the Max-Planck Society and its participating institutes, the Max Planck Institute for Astronomy, Heidelberg, and the Max Planck Institute for Extraterrestrial Physics, Garching, The Johns Hopkins University, Durham University, the University of Edinburgh, the Queen's University Belfast, the Harvard-Smithsonian Center for Astrophysics, the Las Cumbres Observatory Global Telescope Network Incorporated, the National Central University of Taiwan, the Space Telescope Science Institute, the National Aeronautics and Space Administration under Grant No. NNX08AR22G issued through the Planetary Science Division of the NASA Science Mission Directorate, the National Science Foundation Grant No. AST-1238877, the University of Maryland, Eotvos Lorand University (ELTE), the Los Alamos National Laboratory, and the Gordon and Betty Moore Foundation.

This publication makes use of data products from the Two Micron All Sky Survey, which is a joint project of the University of Massachusetts and the Infrared Processing and Analysis Center/California Institute of Technology, funded by the National Aeronautics and Space Administration and the National Science Foundation. This work is based in part on data obtained as part of the UKIRT Infrared Deep Sky Survey.}

\bibliographystyle{apj}
\bibliography{$HOME/Documents/articles/general_desk.bib}

\begin{thebibliography}{}
\expandafter\ifx\csname natexlab\endcsname\relax\def\natexlab#1{#1}\fi

\bibitem[{{Abazajian} {et~al.}(2009){Abazajian}, {Adelman-McCarthy},
  {Ag{\"u}eros}, {Allam}, {Allende Prieto}, {An}, {Anderson}, {Anderson},
  {Annis}, {Bahcall}, {Bailer-Jones}, {Barentine}, {Bassett}, {Becker},
  {Beers}, {Bell}, {Belokurov}, {Berlind}, {Berman}, {Bernardi}, {Bickerton},
  {Bizyaev}, {Blakeslee}, {Blanton}, {Bochanski}, {Boroski}, {Brewington},
  {Brinchmann}, {Brinkmann}, {Brunner}, {Budav{\'a}ri}, {Carey}, {Carliles},
  {Carr}, {Castander}, {Cinabro}, {Connolly}, {Csabai}, {Cunha}, {Czarapata},
  {Davenport}, {de Haas}, {Dilday}, {Doi}, {Eisenstein}, {Evans}, {Evans},
  {Fan}, {Friedman}, {Frieman}, {Fukugita}, {G{\"a}nsicke}, {Gates},
  {Gillespie}, {Gilmore}, {Gonzalez}, {Gonzalez}, {Grebel}, {Gunn},
  {Gy{\"o}ry}, {Hall}, {Harding}, {Harris}, {Harvanek}, {Hawley}, {Hayes},
  {Heckman}, {Hendry}, {Hennessy}, {Hindsley}, {Hoblitt}, {Hogan}, {Hogg},
  {Holtzman}, {Hyde}, {Ichikawa}, {Ichikawa}, {Im}, {Ivezi{\'c}}, {Jester},
  {Jiang}, {Johnson}, {Jorgensen}, {Juri{\'c}}, {Kent}, {Kessler}, {Kleinman},
  {Knapp}, {Konishi}, {Kron}, {Krzesinski}, {Kuropatkin}, {Lampeitl},
  {Lebedeva}, {Lee}, {Lee}, {Leger}, {L{\'e}pine}, {Li}, {Lima}, {Lin}, {Long},
  {Loomis}, {Loveday}, {Lupton}, {Magnier}, {Malanushenko}, {Malanushenko},
  {Mandelbaum}, {Margon}, {Marriner}, {Mart{\'{\i}}nez-Delgado}, {Matsubara},
  {McGehee}, {McKay}, {Meiksin}, {Morrison}, {Mullally}, {Munn}, {Murphy},
  {Nash}, {Nebot}, {Neilsen}, {Newberg}, {Newman}, {Nichol}, {Nicinski},
  {Nieto-Santisteban}, {Nitta}, {Okamura}, {Oravetz}, {Ostriker}, {Owen},
  {Padmanabhan}, {Pan}, {Park}, {Pauls}, {Peoples}, {Percival}, {Pier}, {Pope},
  {Pourbaix}, {Price}, {Purger}, {Quinn}, {Raddick}, {Fiorentin}, {Richards},
  {Richmond}, {Riess}, {Rix}, {Rockosi}, {Sako}, {Schlegel}, {Schneider},
  {Scholz}, {Schreiber}, {Schwope}, {Seljak}, {Sesar}, {Sheldon}, {Shimasaku},
  {Sibley}, {Simmons}, {Sivarani}, {Smith}, {Smith}, {Smol{\v c}i{\'c}},
  {Snedden}, {Stebbins}, {Steinmetz}, {Stoughton}, {Strauss}, {Subba Rao},
  {Suto}, {Szalay}, {Szapudi}, {Szkody}, {Tanaka}, {Tegmark}, {Teodoro},
  {Thakar}, {Tremonti}, {Tucker}, {Uomoto}, {Vanden Berk}, {Vandenberg},
  {Vidrih}, {Vogeley}, {Voges}, {Vogt}, {Wadadekar}, {Watters}, {Weinberg},
  {West}, {White}, {Wilhite}, {Wonders}, {Yanny}, {Yocum}, {York}, {Zehavi},
  {Zibetti}, \& {Zucker}}]{Abazajian09}
{Abazajian}, K.~N., {Adelman-McCarthy}, J.~K., {Ag{\"u}eros}, M.~A., {et~al.}
  2009, \apjs, 182, 543

\bibitem[{{Alexander} {et~al.}(2017){Alexander}, {Wieringa}, {Berger},
  {Saxton}, \& {Komossa}}]{Alexander17}
{Alexander}, K.~D., {Wieringa}, M.~H., {Berger}, E., {Saxton}, R.~D., \&
  {Komossa}, S. 2017, \apj, 837, 153

\bibitem[{{Annis} {et~al.}(2014){Annis}, {Soares-Santos}, {Strauss}, {Becker},
  {Dodelson}, {Fan}, {Gunn}, {Hao}, {Ivezi{\'c}}, {Jester}, {Jiang},
  {Johnston}, {Kubo}, {Lampeitl}, {Lin}, {Lupton}, {Miknaitis}, {Seo}, {Simet},
  \& {Yanny}}]{Annis14}
{Annis}, J., {Soares-Santos}, M., {Strauss}, M.~A., {et~al.} 2014, \apj, 794,
  120

\bibitem[{{Arcavi} {et~al.}(2014){Arcavi}, {Gal-Yam}, {Sullivan}, {Pan},
  {Cenko}, {Horesh}, {Ofek}, {De Cia}, {Yan}, {Yang}, {Howell}, {Tal},
  {Kulkarni}, {Tendulkar}, {Tang}, {Xu}, {Sternberg}, {Cohen}, {Bloom},
  {Nugent}, {Kasliwal}, {Perley}, {Quimby}, {Miller}, {Theissen}, \&
  {Laher}}]{Arcavi14}
{Arcavi}, I., {Gal-Yam}, A., {Sullivan}, M., {et~al.} 2014, \apj, 793, 38

\bibitem[{{Astropy Collaboration} {et~al.}(2013){Astropy Collaboration},
  {Robitaille}, {Tollerud}, {Greenfield}, {Droettboom}, {Bray}, {Aldcroft},
  {Davis}, {Ginsburg}, {Price-Whelan}, {Kerzendorf}, {Conley}, {Crighton},
  {Barbary}, {Muna}, {Ferguson}, {Grollier}, {Parikh}, {Nair}, {Unther},
  {Deil}, {Woillez}, {Conseil}, {Kramer}, {Turner}, {Singer}, {Fox}, {Weaver},
  {Zabalza}, {Edwards}, {Azalee Bostroem}, {Burke}, {Casey}, {Crawford},
  {Dencheva}, {Ely}, {Jenness}, {Labrie}, {Lim}, {Pierfederici}, {Pontzen},
  {Ptak}, {Refsdal}, {Servillat}, \& {Streicher}}]{Astropy-Coll13}
{Astropy Collaboration}, {Robitaille}, T.~P., {Tollerud}, E.~J., {et~al.} 2013,
  \aap, 558, A33

\bibitem[{{Auchettl} {et~al.}(2017{\natexlab{a}}){Auchettl}, {Guillochon}, \&
  {Ramirez-Ruiz}}]{Auchettl16}
{Auchettl}, K., {Guillochon}, J., \& {Ramirez-Ruiz}, E. 2017{\natexlab{a}},
  \apj, 838, 149

\bibitem[{{Auchettl} {et~al.}(2017{\natexlab{b}}){Auchettl}, {Ramirez-Ruiz}, \&
  {Guillochon}}]{Auchettl17}
{Auchettl}, K., {Ramirez-Ruiz}, E., \& {Guillochon}, J. 2017{\natexlab{b}},
  ArXiv e-prints, arXiv:1703.06141

\bibitem[{{Bade} {et~al.}(1996){Bade}, {Komossa}, \& {Dahlem}}]{Bade96}
{Bade}, N., {Komossa}, S., \& {Dahlem}, M. 1996, \aap, 309, L35

\bibitem[{{Baldry} {et~al.}(2012){Baldry}, {Driver}, {Loveday}, {Taylor},
  {Kelvin}, {Liske}, {Norberg}, {Robotham}, {Brough}, {Hopkins}, {Bamford},
  {Peacock}, {Bland-Hawthorn}, {Conselice}, {Croom}, {Jones}, {Parkinson},
  {Popescu}, {Prescott}, {Sharp}, \& {Tuffs}}]{Baldry12}
{Baldry}, I.~K., {Driver}, S.~P., {Loveday}, J., {et~al.} 2012, \mnras, 421,
  621

\bibitem[{{Bertin} {et~al.}(2002){Bertin}, {Ciotti}, \& {Del
  Principe}}]{Bertin02}
{Bertin}, G., {Ciotti}, L., \& {Del Principe}, M. 2002, \aap, 386, 149

\bibitem[{{Bezanson} {et~al.}(2012){Bezanson}, {van Dokkum}, \&
  {Franx}}]{Bezanson12}
{Bezanson}, R., {van Dokkum}, P., \& {Franx}, M. 2012, \apj, 760, 62

\bibitem[{{Bezanson} {et~al.}(2011){Bezanson}, {van Dokkum}, {Franx},
  {Brammer}, {Brinchmann}, {Kriek}, {Labb{\'e}}, {Quadri}, {Rix}, {van de
  Sande}, {Whitaker}, \& {Williams}}]{Bezanson11}
{Bezanson}, R., {van Dokkum}, P.~G., {Franx}, M., {et~al.} 2011, \apjl, 737,
  L31

\bibitem[{{Bielby} {et~al.}(2012){Bielby}, {Hudelot}, {McCracken}, {Ilbert},
  {Daddi}, {Le F{\`e}vre}, {Gonzalez-Perez}, {Kneib}, {Marmo}, {Mellier},
  {Salvato}, {Sanders}, \& {Willott}}]{Bielby12}
{Bielby}, R., {Hudelot}, P., {McCracken}, H.~J., {et~al.} 2012, \aap, 545, A23

\bibitem[{{Blagorodnova} {et~al.}(2017){Blagorodnova}, {Gezari}, {Hung},
  {Kulkarni}, {Cenko}, {Pasham}, {Yan}, {Arcavi}, {Ben-Ami}, {Bue}, {Cantwell},
  {Cao}, {Castro-Tirado}, {Fender}, {Fremling}, {Gal-Yam}, {Ho}, {Horesh},
  {Hosseinzadeh}, {Kasliwal}, {Kong}, {Laher}, {Leloudas}, {Lunnan}, {Masci},
  {Mooley}, {Neill}, {Nugent}, {Powell}, {Valeev}, {Vreeswijk}, {Walters}, \&
  {Wozniak}}]{Blagorodnova17}
{Blagorodnova}, N., {Gezari}, S., {Hung}, T., {et~al.} 2017, \apj, 844, 46

\bibitem[{{Blanchard} {et~al.}(2017){Blanchard}, {Nicholl}, {Berger},
  {Guillochon}, {Margutti}, {Chornock}, {Alexander}, {Leja}, \&
  {Drout}}]{Blanchard17}
{Blanchard}, P.~K., {Nicholl}, M., {Berger}, E., {et~al.} 2017, \apj, 843, 106

\bibitem[{{Blanton} \& {Roweis}(2007)}]{blanton07}
{Blanton}, M.~R., \& {Roweis}, S. 2007, \aj, 133, 734

\bibitem[{{Blanton} {et~al.}(2001){Blanton}, {Dalcanton}, {Eisenstein},
  {Loveday}, {Strauss}, {SubbaRao}, {Weinberg}, {Anderson}, {Annis}, {Bahcall},
  {Bernardi}, {Brinkmann}, {Brunner}, {Burles}, {Carey}, {Castander},
  {Connolly}, {Csabai}, {Doi}, {Finkbeiner}, {Friedman}, {Frieman}, {Fukugita},
  {Gunn}, {Hennessy}, {Hindsley}, {Hogg}, {Ichikawa}, {Ivezi{\'c}}, {Kent},
  {Knapp}, {Lamb}, {Leger}, {Long}, {Lupton}, {McKay}, {Meiksin}, {Merelli},
  {Munn}, {Narayanan}, {Newcomb}, {Nichol}, {Okamura}, {Owen}, {Pier}, {Pope},
  {Postman}, {Quinn}, {Rockosi}, {Schlegel}, {Schneider}, {Shimasaku},
  {Siegmund}, {Smee}, {Snir}, {Stoughton}, {Stubbs}, {Szalay}, {Szokoly},
  {Thakar}, {Tremonti}, {Tucker}, {Uomoto}, {Vanden Berk}, {Vogeley},
  {Waddell}, {Yanny}, {Yasuda}, \& {York}}]{blanton01}
{Blanton}, M.~R., {Dalcanton}, J., {Eisenstein}, D., {et~al.} 2001, \aj, 121,
  2358

\bibitem[{{Blanton} {et~al.}(2005){Blanton}, {Schlegel}, {Strauss},
  {Brinkmann}, {Finkbeiner}, {Fukugita}, {Gunn}, {Hogg}, {Ivezi{\'c}}, {Knapp},
  {Lupton}, {Munn}, {Schneider}, {Tegmark}, \& {Zehavi}}]{Blanton05}
{Blanton}, M.~R., {Schlegel}, D.~J., {Strauss}, M.~A., {et~al.} 2005, \aj, 129,
  2562

\bibitem[{{Bloom} {et~al.}(2011){Bloom}, {Giannios}, {Metzger}, {Cenko},
  {Perley}, {Butler}, {Tanvir}, {Levan}, {O'Brien}, {Strubbe}, {De Colle},
  {Ramirez-Ruiz}, {Lee}, {Nayakshin}, {Quataert}, {King}, {Cucchiara},
  {Guillochon}, {Bower}, {Fruchter}, {Morgan}, \& {van der Horst}}]{Bloom11}
{Bloom}, J.~S., {Giannios}, D., {Metzger}, B.~D., {et~al.} 2011, Science, 333,
  203

\bibitem[{{Brinchmann} {et~al.}(2004){Brinchmann}, {Charlot}, {White},
  {Tremonti}, {Kauffmann}, {Heckman}, \& {Brinkmann}}]{Brinchmann04}
{Brinchmann}, J., {Charlot}, S., {White}, S.~D.~M., {et~al.} 2004, \mnras, 351,
  1151

\bibitem[{{Brockamp} {et~al.}(2011){Brockamp}, {Baumgardt}, \&
  {Kroupa}}]{Brockamp11}
{Brockamp}, M., {Baumgardt}, H., \& {Kroupa}, P. 2011, \mnras, 418, 1308

\bibitem[{{Brown} {et~al.}(2017){Brown}, {Holoien}, {Auchettl}, {Stanek},
  {Kochanek}, {Shappee}, {Prieto}, \& {Grupe}}]{Brown16b}
{Brown}, J.~S., {Holoien}, T.~W.-S., {Auchettl}, K., {et~al.} 2017, \mnras,
  466, 4904

\bibitem[{{Brown} {et~al.}(2016){Brown}, {Shappee}, {W.-S Holoien}, {Stanek},
  {Kochanek}, \& {Prieto}}]{Brown16}
{Brown}, J.~S., {Shappee}, B.~J., {W.-S Holoien}, T., {et~al.} 2016, \mnras,
  462, 3993

\bibitem[{{Brown} {et~al.}(2018){Brown}, {Kochanek}, {Holoien}, {Stanek},
  {Auchettl}, {Shappee}, {Prieto}, {Morrell}, {Falco}, {Strader}, {Chomiuk},
  {Post}, {Villanueva}, {Mathur}, {Dong}, {Chen}, \& {Bose}}]{Brown18}
{Brown}, J.~S., {Kochanek}, C.~S., {Holoien}, T.~W.-S., {et~al.} 2018, \mnras,
  473, 1130

\bibitem[{{Cenko} {et~al.}(2012){Cenko}, {Bloom}, {Kulkarni}, {Strubbe},
  {Miller}, {Butler}, {Quimby}, {Gal-Yam}, {Ofek}, {Quataert}, {Bildsten},
  {Poznanski}, {Perley}, {Morgan}, {Filippenko}, {Frail}, {Arcavi}, {Ben-Ami},
  {Cucchiara}, {Fassnacht}, {Green}, {Hook}, {Howell}, {Lagattuta}, {Law},
  {Kasliwal}, {Nugent}, {Silverman}, {Sullivan}, {Tendulkar}, \&
  {Yaron}}]{Cenko12}
{Cenko}, S.~B., {Bloom}, J.~S., {Kulkarni}, S.~R., {et~al.} 2012, \mnras, 420,
  2684

\bibitem[{{Chambers}(2007)}]{Chambers07}
{Chambers}, K.~C. 2007, in Bulletin of the AAS, Vol.~38, 995

\bibitem[{{Chambers} {et~al.}(2016){Chambers}, {Magnier}, {Metcalfe},
  {Flewelling}, {Huber}, {Waters}, {Denneau}, {Draper}, {Farrow}, {Finkbeiner},
  {Holmberg}, {Koppenhoefer}, {Price}, {Saglia}, {Schlafly}, {Smartt},
  {Sweeney}, {Wainscoat}, {Burgett}, {Grav}, {Heasley}, {Hodapp}, {Jedicke},
  {Kaiser}, {Kudritzki}, {Luppino}, {Lupton}, {Monet}, {Morgan}, {Onaka},
  {Stubbs}, {Tonry}, {Banados}, {Bell}, {Bender}, {Bernard}, {Botticella},
  {Casertano}, {Chastel}, {Chen}, {Chen}, {Cole}, {Deacon}, {Frenk},
  {Fitzsimmons}, {Gezari}, {Goessl}, {Goggia}, {Goldman}, {Grebel}, {Hambly},
  {Hasinger}, {Heavens}, {Heckman}, {Henderson}, {Henning}, {Holman}, {Hopp},
  {Ip}, {Isani}, {Keyes}, {Koekemoer}, {Kotak}, {Long}, {Lucey}, {Liu},
  {Martin}, {McLean}, {Morganson}, {Murphy}, {Nieto-Santisteban}, {Norberg},
  {Peacock}, {Pier}, {Postman}, {Primak}, {Rae}, {Rest}, {Riess}, {Riffeser},
  {Rix}, {Roser}, {Schilbach}, {Schultz}, {Scolnic}, {Szalay}, {Seitz},
  {Shiao}, {Small}, {Smith}, {Soderblom}, {Taylor}, {Thakar}, {Thiel},
  {Thilker}, {Urata}, {Valenti}, {Walter}, {Watters}, {Werner}, {White},
  {Wood-Vasey}, \& {Wyse}}]{Chambers16}
{Chambers}, K.~C., {Magnier}, E.~A., {Metcalfe}, N., {et~al.} 2016, ArXiv
  e-prints, arXiv:1612.05560

\bibitem[{{Chornock} {et~al.}(2014){Chornock}, {Berger}, {Gezari}, {Zauderer},
  {Rest}, {Chomiuk}, {Kamble}, {Soderberg}, {Czekala}, {Dittmann}, {Drout},
  {Foley}, {Fong}, {Huber}, {Kirshner}, {Lawrence}, {Lunnan}, {Marion},
  {Narayan}, {Riess}, {Roth}, {Sanders}, {Scolnic}, {Smartt}, {Smith},
  {Stubbs}, {Tonry}, {Burgett}, {Chambers}, {Flewelling}, {Hodapp}, {Kaiser},
  {Magnier}, {Martin}, {Neill}, {Price}, \& {Wainscoat}}]{Chornock14}
{Chornock}, R., {Berger}, E., {Gezari}, S., {et~al.} 2014, \apj, 780, 44

\bibitem[{{Cool} {et~al.}(2012){Cool}, {Eisenstein}, {Kochanek}, {Brown},
  {Caldwell}, {Dey}, {Forman}, {Hickox}, {Jannuzi}, {Jones}, {Moustakas}, \&
  {Murray}}]{Cool12}
{Cool}, R.~J., {Eisenstein}, D.~J., {Kochanek}, C.~S., {et~al.} 2012, \apj,
  748, 10

\bibitem[{{Czerny} {et~al.}(2009){Czerny}, {Siemiginowska}, {Janiuk},
  {Nikiel-Wroczy{\'n}ski}, \& {Stawarz}}]{Czerny09}
{Czerny}, B., {Siemiginowska}, A., {Janiuk}, A., {Nikiel-Wroczy{\'n}ski}, B.,
  \& {Stawarz}, {\L}. 2009, \apj, 698, 840

\bibitem[{{Dai} {et~al.}(2015){Dai}, {McKinney}, \& {Miller}}]{Dai15}
{Dai}, L., {McKinney}, J.~C., \& {Miller}, M.~C. 2015, \apjl, 812, L39

\bibitem[{{Dong} {et~al.}(2016){Dong}, {Shappee}, {Prieto}, {Jha}, {Stanek},
  {Holoien}, {Kochanek}, {Thompson}, {Morrell}, {Thompson}, {Basu}, {Beacom},
  {Bersier}, {Brimacombe}, {Brown}, {Bufano}, {Chen}, {Conseil}, {Danilet},
  {Falco}, {Grupe}, {Kiyota}, {Masi}, {Nicholls}, {Olivares E.}, {Pignata},
  {Pojmanski}, {Simonian}, {Szczygiel}, \& {Wo{\'z}niak}}]{Dong16}
{Dong}, S., {Shappee}, B.~J., {Prieto}, J.~L., {et~al.} 2016, Science, 351, 257

\bibitem[{{Drake} {et~al.}(2011){Drake}, {Djorgovski}, {Mahabal}, {Anderson},
  {Roy}, {Mohan}, {Ravindranath}, {Frail}, {Gezari}, {Neill}, {Ho}, {Prieto},
  {Thompson}, {Thorstensen}, {Wagner}, {Kowalski}, {Chiang}, {Grove},
  {Schinzel}, {Wood}, {Carrasco}, {Recillas}, {Kewley}, {Archana}, {Basu},
  {Wadadekar}, {Kumar}, {Myers}, {Phinney}, {Williams}, {Graham}, {Catelan},
  {Beshore}, {Larson}, \& {Christensen}}]{Drake11}
{Drake}, A.~J., {Djorgovski}, S.~G., {Mahabal}, A., {et~al.} 2011, \apj, 735,
  106

\bibitem[{{Dressler} \& {Gunn}(1983)}]{Dressler83}
{Dressler}, A., \& {Gunn}, J.~E. 1983, \apj, 270, 7

\bibitem[{{Ferrarese} \& {Merritt}(2000)}]{Ferrarese00}
{Ferrarese}, L., \& {Merritt}, D. 2000, \apjl, 539, L9

\bibitem[{{Flewelling} {et~al.}(2016){Flewelling}, {Magnier}, {Chambers},
  {Heasley}, {Holmberg}, {Huber}, {Sweeney}, {Waters}, {Chen}, {Farrow},
  {Hasinger}, {Henderson}, {Long}, {Metcalfe}, {Nieto-Santisteban}, {Norberg},
  {Saglia}, {Szalay}, {Rest}, {Thakar}, {Tonry}, {Valenti}, {Werner}, {White},
  {Denneau}, {Draper}, {Hodapp}, {Jedicke}, {Kaiser}, {Kudritzki}, {Price},
  {Wainscoat}, {Chastel}, {McClean}, {Postman}, \& {Shiao}}]{Flewelling16}
{Flewelling}, H.~A., {Magnier}, E.~A., {Chambers}, K.~C., {et~al.} 2016, ArXiv
  e-prints, arXiv:1612.05243

\bibitem[{{French} {et~al.}(2016){French}, {Arcavi}, \& {Zabludoff}}]{French16}
{French}, K.~D., {Arcavi}, I., \& {Zabludoff}, A. 2016, \apjl, 818, L21

\bibitem[{{French} {et~al.}(2017){French}, {Arcavi}, \& {Zabludoff}}]{French17}
---. 2017, \apj, 835, 176

\bibitem[{{Frieman} {et~al.}(2008){Frieman}, {Bassett}, {Becker}, {Choi},
  {Cinabro}, {DeJongh}, {Depoy}, {Dilday}, {Doi}, {Garnavich}, {Hogan},
  {Holtzman}, {Im}, {Jha}, {Kessler}, {Konishi}, {Lampeitl}, {Marriner},
  {Marshall}, {McGinnis}, {Miknaitis}, {Nichol}, {Prieto}, {Riess}, {Richmond},
  {Romani}, {Sako}, {Schneider}, {Smith}, {Takanashi}, {Tokita}, {van der
  Heyden}, {Yasuda}, {Zheng}, {Adelman-McCarthy}, {Annis}, {Assef},
  {Barentine}, {Bender}, {Blandford}, {Boroski}, {Bremer}, {Brewington},
  {Collins}, {Crotts}, {Dembicky}, {Eastman}, {Edge}, {Edmondson}, {Elson},
  {Eyler}, {Filippenko}, {Foley}, {Frank}, {Goobar}, {Gueth}, {Gunn},
  {Harvanek}, {Hopp}, {Ihara}, {Ivezi{\'c}}, {Kahn}, {Kaplan}, {Kent},
  {Ketzeback}, {Kleinman}, {Kollatschny}, {Kron}, {Krzesi{\'n}ski}, {Lamenti},
  {Leloudas}, {Lin}, {Long}, {Lucey}, {Lupton}, {Malanushenko}, {Malanushenko},
  {McMillan}, {Mendez}, {Morgan}, {Morokuma}, {Nitta}, {Ostman}, {Pan},
  {Rockosi}, {Romer}, {Ruiz-Lapuente}, {Saurage}, {Schlesinger}, {Snedden},
  {Sollerman}, {Stoughton}, {Stritzinger}, {Subba Rao}, {Tucker}, {Vaisanen},
  {Watson}, {Watters}, {Wheeler}, {Yanny}, \& {York}}]{frieman08}
{Frieman}, J.~A., {Bassett}, B., {Becker}, A., {et~al.} 2008, \aj, 135, 338

\bibitem[{{Fruchter} {et~al.}(2006){Fruchter}, {Levan}, {Strolger},
  {Vreeswijk}, {Thorsett}, {Bersier}, {Burud}, {Castro Cer{\'o}n},
  {Castro-Tirado}, {Conselice}, {Dahlen}, {Ferguson}, {Fynbo}, {Garnavich},
  {Gibbons}, {Gorosabel}, {Gull}, {Hjorth}, {Holland}, {Kouveliotou}, {Levay},
  {Livio}, {Metzger}, {Nugent}, {Petro}, {Pian}, {Rhoads}, {Riess}, {Sahu},
  {Smette}, {Tanvir}, {Wijers}, \& {Woosley}}]{Fruchter06}
{Fruchter}, A.~S., {Levan}, A.~J., {Strolger}, L., {et~al.} 2006, \nat, 441,
  463

\bibitem[{{Gebhardt} {et~al.}(2000){Gebhardt}, {Bender}, {Bower}, {Dressler},
  {Faber}, {Filippenko}, {Green}, {Grillmair}, {Ho}, {Kormendy}, {Lauer},
  {Magorrian}, {Pinkney}, {Richstone}, \& {Tremaine}}]{Gebhardt00}
{Gebhardt}, K., {Bender}, R., {Bower}, G., {et~al.} 2000, \apjl, 539, L13

\bibitem[{{Gezari} {et~al.}(2015){Gezari}, {Chornock}, {Lawrence}, {Rest},
  {Jones}, {Berger}, {Challis}, \& {Narayan}}]{Gezari15}
{Gezari}, S., {Chornock}, R., {Lawrence}, A., {et~al.} 2015, \apjl, 815, L5

\bibitem[{{Gezari} {et~al.}(2006){Gezari}, {Martin}, {Milliard}, {Basa},
  {Halpern}, {Forster}, {Friedman}, {Morrissey}, {Neff}, {Schiminovich},
  {Seibert}, {Small}, \& {Wyder}}]{Gezari06}
{Gezari}, S., {Martin}, D.~C., {Milliard}, B., {et~al.} 2006, \apjl, 653, L25

\bibitem[{{Gezari} {et~al.}(2008){Gezari}, {Basa}, {Martin}, {Bazin},
  {Forster}, {Milliard}, {Halpern}, {Friedman}, {Morrissey}, {Neff},
  {Schiminovich}, {Seibert}, {Small}, \& {Wyder}}]{Gezari08}
{Gezari}, S., {Basa}, S., {Martin}, D.~C., {et~al.} 2008, \apj, 676, 944

\bibitem[{{Gezari} {et~al.}(2009){Gezari}, {Heckman}, {Cenko}, {Eracleous},
  {Forster}, {Gon{\c c}alves}, {Martin}, {Morrissey}, {Neff}, {Seibert},
  {Schiminovich}, \& {Wyder}}]{Gezari09}
{Gezari}, S., {Heckman}, T., {Cenko}, S.~B., {et~al.} 2009, \apj, 698, 1367

\bibitem[{{Gezari} {et~al.}(2012){Gezari}, {Chornock}, {Rest}, {Huber},
  {Forster}, {Berger}, {Challis}, {Neill}, {Martin}, {Heckman}, {Lawrence},
  {Norman}, {Narayan}, {Foley}, {Marion}, {Scolnic}, {Chomiuk}, {Soderberg},
  {Smith}, {Kirshner}, {Riess}, {Smartt}, {Stubbs}, {Tonry}, {Wood-Vasey},
  {Burgett}, {Chambers}, {Grav}, {Heasley}, {Kaiser}, {Kudritzki}, {Magnier},
  {Morgan}, \& {Price}}]{Gezari12}
{Gezari}, S., {Chornock}, R., {Rest}, A., {et~al.} 2012, \nat, 485, 217

\bibitem[{{Gezari} {et~al.}(2017){Gezari}, {Hung}, {Cenko}, {Blagorodnova},
  {Yan}, {Kulkarni}, {Mooley}, {Kong}, {Cantwell}, {Yu}, {Cao}, {Fremling},
  {Neill}, {Ngeow}, {Nugent}, \& {Wozniak}}]{Gezari17}
{Gezari}, S., {Hung}, T., {Cenko}, S.~B., {et~al.} 2017, \apj, 835, 144

\bibitem[{{Godoy-Rivera} {et~al.}(2017){Godoy-Rivera}, {Stanek}, {Kochanek},
  {Chen}, {Dong}, {Prieto}, {Shappee}, {Jha}, {Foley}, {Pan}, {Holoien},
  {Thompson}, {Grupe}, \& {Beacom}}]{Godoy-Rivera17}
{Godoy-Rivera}, D., {Stanek}, K.~Z., {Kochanek}, C.~S., {et~al.} 2017, \mnras,
  466, 1428

\bibitem[{{Graham} {et~al.}(2017){Graham}, {Djorgovski}, {Drake}, {Stern},
  {Mahabal}, {Glikman}, {Larson}, \& {Christensen}}]{Graham17}
{Graham}, M.~J., {Djorgovski}, S.~G., {Drake}, A.~J., {et~al.} 2017, \mnras,
  470, 4112

\bibitem[{{Graur} {et~al.}(2017){Graur}, {French}, {Zahid}, {Guillochon},
  {Mandel}, {Auchettl}, \& {Zabludoff}}]{Graur17}
{Graur}, O., {French}, K.~D., {Zahid}, H.~J., {et~al.} 2017, ArXiv e-prints,
  arXiv:1707.02986

\bibitem[{{Grogin} {et~al.}(2011){Grogin}, {Kocevski}, {Faber}, {Ferguson},
  {Koekemoer}, {Riess}, {Acquaviva}, {Alexander}, {Almaini}, {Ashby}, {Barden},
  {Bell}, {Bournaud}, {Brown}, {Caputi}, {Casertano}, {Cassata}, {Castellano},
  {Challis}, {Chary}, {Cheung}, {Cirasuolo}, {Conselice}, {Roshan Cooray},
  {Croton}, {Daddi}, {Dahlen}, {Dav{\'e}}, {de Mello}, {Dekel}, {Dickinson},
  {Dolch}, {Donley}, {Dunlop}, {Dutton}, {Elbaz}, {Fazio}, {Filippenko},
  {Finkelstein}, {Fontana}, {Gardner}, {Garnavich}, {Gawiser}, {Giavalisco},
  {Grazian}, {Guo}, {Hathi}, {H{\"a}ussler}, {Hopkins}, {Huang}, {Huang},
  {Jha}, {Kartaltepe}, {Kirshner}, {Koo}, {Lai}, {Lee}, {Li}, {Lotz}, {Lucas},
  {Madau}, {McCarthy}, {McGrath}, {McIntosh}, {McLure}, {Mobasher},
  {Moustakas}, {Mozena}, {Nandra}, {Newman}, {Niemi}, {Noeske}, {Papovich},
  {Pentericci}, {Pope}, {Primack}, {Rajan}, {Ravindranath}, {Reddy}, {Renzini},
  {Rix}, {Robaina}, {Rodney}, {Rosario}, {Rosati}, {Salimbeni}, {Scarlata},
  {Siana}, {Simard}, {Smidt}, {Somerville}, {Spinrad}, {Straughn}, {Strolger},
  {Telford}, {Teplitz}, {Trump}, {van der Wel}, {Villforth}, {Wechsler},
  {Weiner}, {Wiklind}, {Wild}, {Wilson}, {Wuyts}, {Yan}, \& {Yun}}]{Grogin11}
{Grogin}, N.~A., {Kocevski}, D.~D., {Faber}, S.~M., {et~al.} 2011, \apjs, 197,
  35

\bibitem[{{Guillochon} {et~al.}(2014){Guillochon}, {Manukian}, \&
  {Ramirez-Ruiz}}]{Guillochon14}
{Guillochon}, J., {Manukian}, H., \& {Ramirez-Ruiz}, E. 2014, \apj, 783, 23

\bibitem[{{Guillochon} \& {Ramirez-Ruiz}(2013)}]{Guillochon13}
{Guillochon}, J., \& {Ramirez-Ruiz}, E. 2013, \apj, 767, 25

\bibitem[{{Guillochon} \& {Ramirez-Ruiz}(2015)}]{Guillochon15}
---. 2015, \apj, 809, 166

\bibitem[{{G{\"u}ltekin} {et~al.}(2009){G{\"u}ltekin}, {Richstone}, {Gebhardt},
  {Lauer}, {Tremaine}, {Aller}, {Bender}, {Dressler}, {Faber}, {Filippenko},
  {Green}, {Ho}, {Kormendy}, {Magorrian}, {Pinkney}, \& {Siopis}}]{Gultekin09}
{G{\"u}ltekin}, K., {Richstone}, D.~O., {Gebhardt}, K., {et~al.} 2009, \apj,
  698, 198

\bibitem[{{Hambly} {et~al.}(2008){Hambly}, {Collins}, {Cross}, {Mann}, {Read},
  {Sutorius}, {Bond}, {Bryant}, {Emerson}, {Lawrence}, {Rimoldini}, {Stewart},
  {Williams}, {Adamson}, {Hirst}, {Dye}, \& {Warren}}]{Hambly08}
{Hambly}, N.~C., {Collins}, R.~S., {Cross}, N.~J.~G., {et~al.} 2008, \mnras,
  384, 637

\bibitem[{{Hameury} {et~al.}(2009){Hameury}, {Viallet}, \&
  {Lasota}}]{Hameury09}
{Hameury}, J.-M., {Viallet}, M., \& {Lasota}, J.-P. 2009, \aap, 496, 413

\bibitem[{{Hatziminaoglou} {et~al.}(2001){Hatziminaoglou}, {Siemiginowska}, \&
  {Elvis}}]{Hatziminaogl01}
{Hatziminaoglou}, E., {Siemiginowska}, A., \& {Elvis}, M. 2001, \apj, 547, 90

\bibitem[{{Heckman}(1980)}]{heckman80}
{Heckman}, T.~M. 1980, \aap, 87, 152

\bibitem[{{Hills}(1975)}]{Hills75}
{Hills}, J.~G. 1975, \nat, 254, 295

\bibitem[{{Hogg} {et~al.}(2002){Hogg}, {Baldry}, {Blanton}, \&
  {Eisenstein}}]{Hogg02}
{Hogg}, D.~W., {Baldry}, I.~K., {Blanton}, M.~R., \& {Eisenstein}, D.~J. 2002,
  ArXiv astro-ph/0210394, astro-ph/0210394

\bibitem[{{Holoien} {et~al.}(2014){Holoien}, {Prieto}, {Bersier}, {Kochanek},
  {Stanek}, {Shappee}, {Grupe}, {Basu}, {Beacom}, {Brimacombe}, {Brown},
  {Davis}, {Jencson}, {Pojmanski}, \& {Szczygie{\l}}}]{Holoien14}
{Holoien}, T.~W.-S., {Prieto}, J.~L., {Bersier}, D., {et~al.} 2014, \mnras,
  445, 3263

\bibitem[{{Holoien} {et~al.}(2016{\natexlab{a}}){Holoien}, {Kochanek},
  {Prieto}, {Grupe}, {Chen}, {Godoy-Rivera}, {Stanek}, {Shappee}, {Dong},
  {Brown}, {Basu}, {Beacom}, {Bersier}, {Brimacombe}, {Carlson}, {Falco},
  {Johnston}, {Madore}, {Pojmanski}, \& {Seibert}}]{Holoien16b}
{Holoien}, T.~W.-S., {Kochanek}, C.~S., {Prieto}, J.~L., {et~al.}
  2016{\natexlab{a}}, \mnras, 463, 3813

\bibitem[{{Holoien} {et~al.}(2016{\natexlab{b}}){Holoien}, {Kochanek},
  {Prieto}, {Stanek}, {Dong}, {Shappee}, {Grupe}, {Brown}, {Basu}, {Beacom},
  {Bersier}, {Brimacombe}, {Danilet}, {Falco}, {Guo}, {Jose}, {Herczeg},
  {Long}, {Pojmanski}, {Simonian}, {Szczygie{\l}}, {Thompson}, {Thorstensen},
  {Wagner}, \& {Wo{\'z}niak}}]{Holoien16}
---. 2016{\natexlab{b}}, \mnras, 455, 2918

\bibitem[{{Hung} {et~al.}(2017){Hung}, {Gezari}, {Blagorodnova}, {Roth},
  {Cenko}, {Kulkarni}, {Horesh}, {Arcavi}, {McCully}, {Yan}, {Lunnan},
  {Fremling}, {Cao}, {Nugent}, \& {Wozniak}}]{Hung17}
{Hung}, T., {Gezari}, S., {Blagorodnova}, N., {et~al.} 2017, \apj, 842, 29

\bibitem[{{Janiuk} {et~al.}(2002){Janiuk}, {Czerny}, \&
  {Siemiginowska}}]{Janiuk02}
{Janiuk}, A., {Czerny}, B., \& {Siemiginowska}, A. 2002, \apj, 576, 908

\bibitem[{{Jarrett} {et~al.}(2000){Jarrett}, {Chester}, {Cutri}, {Schneider},
  {Skrutskie}, \& {Huchra}}]{Jarrett00}
{Jarrett}, T.~H., {Chester}, T., {Cutri}, R., {et~al.} 2000, \aj, 119, 2498

\bibitem[{{Jiang} {et~al.}(2016){Jiang}, {Dou}, {Wang}, {Yang}, {Lyu}, \&
  {Zhou}}]{Jiang16}
{Jiang}, N., {Dou}, L., {Wang}, T., {et~al.} 2016, \apjl, 828, L14

\bibitem[{{Kauffmann} {et~al.}(2003){Kauffmann}, {Heckman}, {White}, {Charlot},
  {Tremonti}, {Brinchmann}, {Bruzual}, {Peng}, {Seibert}, {Bernardi},
  {Blanton}, {Brinkmann}, {Castander}, {Cs{\'a}bai}, {Fukugita}, {Ivezic},
  {Munn}, {Nichol}, {Padmanabhan}, {Thakar}, {Weinberg}, \&
  {York}}]{Kauffmann03b}
{Kauffmann}, G., {Heckman}, T.~M., {White}, S.~D.~M., {et~al.} 2003, \mnras,
  341, 33

\bibitem[{{Kaviraj} {et~al.}(2007){Kaviraj}, {Kirkby}, {Silk}, \&
  {Sarzi}}]{Kaviraj07}
{Kaviraj}, S., {Kirkby}, L.~A., {Silk}, J., \& {Sarzi}, M. 2007, \mnras, 382,
  960

\bibitem[{{Kesden}(2012)}]{Kesden12b}
{Kesden}, M. 2012, \prd, 85, 024037

\bibitem[{{Kochanek}(2016)}]{Kochanek16}
{Kochanek}, C.~S. 2016, \mnras, 461, 371

\bibitem[{{Koekemoer} {et~al.}(2011){Koekemoer}, {Faber}, {Ferguson}, {Grogin},
  {Kocevski}, {Koo}, {Lai}, {Lotz}, {Lucas}, {McGrath}, {Ogaz}, {Rajan},
  {Riess}, {Rodney}, {Strolger}, {Casertano}, {Castellano}, {Dahlen},
  {Dickinson}, {Dolch}, {Fontana}, {Giavalisco}, {Grazian}, {Guo}, {Hathi},
  {Huang}, {van der Wel}, {Yan}, {Acquaviva}, {Alexander}, {Almaini}, {Ashby},
  {Barden}, {Bell}, {Bournaud}, {Brown}, {Caputi}, {Cassata}, {Challis},
  {Chary}, {Cheung}, {Cirasuolo}, {Conselice}, {Roshan Cooray}, {Croton},
  {Daddi}, {Dav{\'e}}, {de Mello}, {de Ravel}, {Dekel}, {Donley}, {Dunlop},
  {Dutton}, {Elbaz}, {Fazio}, {Filippenko}, {Finkelstein}, {Frazer}, {Gardner},
  {Garnavich}, {Gawiser}, {Gruetzbauch}, {Hartley}, {H{\"a}ussler},
  {Herrington}, {Hopkins}, {Huang}, {Jha}, {Johnson}, {Kartaltepe},
  {Khostovan}, {Kirshner}, {Lani}, {Lee}, {Li}, {Madau}, {McCarthy},
  {McIntosh}, {McLure}, {McPartland}, {Mobasher}, {Moreira}, {Mortlock},
  {Moustakas}, {Mozena}, {Nandra}, {Newman}, {Nielsen}, {Niemi}, {Noeske},
  {Papovich}, {Pentericci}, {Pope}, {Primack}, {Ravindranath}, {Reddy},
  {Renzini}, {Rix}, {Robaina}, {Rosario}, {Rosati}, {Salimbeni}, {Scarlata},
  {Siana}, {Simard}, {Smidt}, {Snyder}, {Somerville}, {Spinrad}, {Straughn},
  {Telford}, {Teplitz}, {Trump}, {Vargas}, {Villforth}, {Wagner}, {Wandro},
  {Wechsler}, {Weiner}, {Wiklind}, {Wild}, {Wilson}, {Wuyts}, \&
  {Yun}}]{Koekemoer11}
{Koekemoer}, A.~M., {Faber}, S.~M., {Ferguson}, H.~C., {et~al.} 2011, \apjs,
  197, 36

\bibitem[{{Komossa}(2015)}]{Komossa15}
{Komossa}, S. 2015, Journal of High Energy Astrophysics, 7, 148

\bibitem[{{Komossa} \& {Bade}(1999)}]{KomossaBade99}
{Komossa}, S., \& {Bade}, N. 1999, \aap, 343, 775

\bibitem[{{Komossa} {et~al.}(2008){Komossa}, {Zhou}, {Wang}, {Ajello}, {Ge},
  {Greiner}, {Lu}, {Salvato}, {Saxton}, {Shan}, {Xu}, \& {Yuan}}]{Komossa08}
{Komossa}, S., {Zhou}, H., {Wang}, T., {et~al.} 2008, \apjl, 678, L13

\bibitem[{{Krolik} {et~al.}(2016){Krolik}, {Piran}, {Svirski}, \&
  {Cheng}}]{Krolik16}
{Krolik}, J., {Piran}, T., {Svirski}, G., \& {Cheng}, R.~M. 2016, \apj, 827,
  127

\bibitem[{{Kroupa} {et~al.}(1993){Kroupa}, {Tout}, \& {Gilmore}}]{Kroupa93}
{Kroupa}, P., {Tout}, C.~A., \& {Gilmore}, G. 1993, \mnras, 262, 545

\bibitem[{{Kr{\"u}hler} {et~al.}(2017){Kr{\"u}hler}, {Fraser}, {Leloudas},
  {Schulze}, {Stone}, {van Velzen}, {Amorin}, {Hjorth}, {Jonker}, {Kann},
  {Kim}, {Kuncarayakti}, {Mehner}, \& {Nicuesa Guelbenzu}}]{Kruhler17}
{Kr{\"u}hler}, T., {Fraser}, M., {Leloudas}, G., {et~al.} 2017, ArXiv e-prints,
  arXiv:1710.01045

\bibitem[{{Lackner} \& {Gunn}(2012)}]{Lackner12}
{Lackner}, C.~N., \& {Gunn}, J.~E. 2012, \mnras, 421, 2277

\bibitem[{{LaMassa} {et~al.}(2015){LaMassa}, {Cales}, {Moran}, {Myers},
  {Richards}, {Eracleous}, {Heckman}, {Gallo}, \& {Urry}}]{LaMassa15}
{LaMassa}, S.~M., {Cales}, S., {Moran}, E.~C., {et~al.} 2015, \apj, 800, 144

\bibitem[{{Lang} {et~al.}(2014){Lang}, {Hogg}, \& {Schlegel}}]{Lang14b}
{Lang}, D., {Hogg}, D.~W., \& {Schlegel}, D.~J. 2014, ArXiv e-prints,
  arXiv:1410.7397

\bibitem[{{Lasota}(2001)}]{Lasota01}
{Lasota}, J.-P. 2001, New A Rev., 45, 449

\bibitem[{{Law} {et~al.}(2009){Law}, {Kulkarni}, {Dekany}, {Ofek}, {Quimby},
  {Nugent}, {Surace}, {Grillmair}, {Bloom}, {Kasliwal}, {Bildsten}, {Brown},
  {Cenko}, {Ciardi}, {Croner}, {Djorgovski}, {van Eyken}, {Filippenko}, {Fox},
  {Gal-Yam}, {Hale}, {Hamam}, {Helou}, {Henning}, {Howell}, {Jacobsen},
  {Laher}, {Mattingly}, {McKenna}, {Pickles}, {Poznanski}, {Rahmer}, {Rau},
  {Rosing}, {Shara}, {Smith}, {Starr}, {Sullivan}, {Velur}, {Walters}, \&
  {Zolkower}}]{Law09}
{Law}, N.~M., {Kulkarni}, S.~R., {Dekany}, R.~G., {et~al.} 2009, \pasp, 121,
  1395

\bibitem[{{Law-Smith} {et~al.}(2017){Law-Smith}, {Ramirez-Ruiz}, {Ellison}, \&
  {Foley}}]{Law-Smith17}
{Law-Smith}, J., {Ramirez-Ruiz}, E., {Ellison}, S.~L., \& {Foley}, R.~J. 2017,
  \apj, 850, 22

\bibitem[{{Lawrence} {et~al.}(2007){Lawrence}, {Warren}, {Almaini}, {Edge},
  {Hambly}, {Jameson}, {Lucas}, {Casali}, {Adamson}, {Dye}, {Emerson},
  {Foucaud}, {Hewett}, {Hirst}, {Hodgkin}, {Irwin}, {Lodieu}, {McMahon},
  {Simpson}, {Smail}, {Mortlock}, \& {Folger}}]{Lawrence07}
{Lawrence}, A., {Warren}, S.~J., {Almaini}, O., {et~al.} 2007, \mnras, 379,
  1599

\bibitem[{{Leloudas} {et~al.}(2016){Leloudas}, {Fraser}, {Stone}, {van Velzen},
  {Jonker}, {Arcavi}, {Fremling}, {Maund}, {Smartt}, {Kr{\`\i}hler},
  {Miller-Jones}, {Vreeswijk}, {Gal-Yam}, {Mazzali}, {De Cia}, {Howell},
  {Inserra}, {Patat}, {de Ugarte Postigo}, {Yaron}, {Ashall}, {Bar},
  {Campbell}, {Chen}, {Childress}, {Elias-Rosa}, {Harmanen}, {Hosseinzadeh},
  {Johansson}, {Kangas}, {Kankare}, {Kim}, {Kuncarayakti}, {Lyman}, {Magee},
  {Maguire}, {Malesani}, {Mattila}, {McCully}, {Nicholl}, {Prentice},
  {Romero-Ca{\~n}izales}, {Schulze}, {Smith}, {Sollerman}, {Sullivan},
  {Tucker}, {Valenti}, {Wheeler}, \& {Young}}]{Leloudas16}
{Leloudas}, G., {Fraser}, M., {Stone}, N.~C., {et~al.} 2016, Nature Astronomy,
  1, 0002

\bibitem[{{Levan} {et~al.}(2011){Levan}, {Tanvir}, {Cenko}, {Perley},
  {Wiersema}, {Bloom}, {Fruchter}, {Postigo}, {O'Brien}, {Butler}, {van der
  Horst}, {Leloudas}, {Morgan}, {Misra}, {Bower}, {Farihi}, {Tunnicliffe},
  {Modjaz}, {Silverman}, {Hjorth}, {Th{\"o}ne}, {Cucchiara}, {Cer{\'o}n},
  {Castro-Tirado}, {Arnold}, {Bremer}, {Brodie}, {Carroll}, {Cooper}, {Curran},
  {Cutri}, {Ehle}, {Forbes}, {Fynbo}, {Gorosabel}, {Graham}, {Hoffman},
  {Guziy}, {Jakobsson}, {Kamble}, {Kerr}, {Kasliwal}, {Kouveliotou},
  {Kocevski}, {Law}, {Nugent}, {Ofek}, {Poznanski}, {Quimby}, {Rol},
  {Romanowsky}, {S{\'a}nchez-Ram{\'{\i}}rez}, {Schulze}, {Singh}, {van
  Spaandonk}, {Starling}, {Strom}, {Tello}, {Vaduvescu}, {Wheatley}, {Wijers},
  {Winters}, \& {Xu}}]{Levan11}
{Levan}, A.~J., {Tanvir}, N.~R., {Cenko}, S.~B., {et~al.} 2011, Science, 333,
  199

\bibitem[{{Lodato} \& {Rossi}(2011)}]{Lodato11}
{Lodato}, G., \& {Rossi}, E.~M. 2011, \mnras, 410, 359

\bibitem[{{Lu} {et~al.}(2017){Lu}, {Kumar}, \& {Narayan}}]{Lu17}
{Lu}, W., {Kumar}, P., \& {Narayan}, R. 2017, \mnras, 468, 910

\bibitem[{{MacLeod} {et~al.}(2012){MacLeod}, {Ivezi{\'c}}, {Sesar}, {de Vries},
  {Kochanek}, {Kelly}, {Becker}, {Lupton}, {Hall}, {Richards}, {Anderson}, \&
  {Schneider}}]{MacLeod12}
{MacLeod}, C.~L., {Ivezi{\'c}}, {\v Z}., {Sesar}, B., {et~al.} 2012, \apj, 753,
  106

\bibitem[{{MacLeod} {et~al.}(2016){MacLeod}, {Ross}, {Lawrence}, {Goad},
  {Horne}, {Burgett}, {Chambers}, {Flewelling}, {Hodapp}, {Kaiser}, {Magnier},
  {Wainscoat}, \& {Waters}}]{MacLeod16}
{MacLeod}, C.~L., {Ross}, N.~P., {Lawrence}, A., {et~al.} 2016, \mnras, 457,
  389

\bibitem[{{MacLeod} {et~al.}(2014){MacLeod}, {Goldstein}, {Ramirez-Ruiz},
  {Guillochon}, \& {Samsing}}]{MacLeod14}
{MacLeod}, M., {Goldstein}, J., {Ramirez-Ruiz}, E., {Guillochon}, J., \&
  {Samsing}, J. 2014, \apj, 794, 9

\bibitem[{{Mainzer} {et~al.}(2014){Mainzer}, {Bauer}, {Cutri}, {Grav},
  {Masiero}, {Beck}, {Clarkson}, {Conrow}, {Dailey}, {Eisenhardt}, {Fabinsky},
  {Fajardo-Acosta}, {Fowler}, {Gelino}, {Grillmair}, {Heinrichsen}, {Kendall},
  {Kirkpatrick}, {Liu}, {Masci}, {McCallon}, {Nugent}, {Papin}, {Rice},
  {Royer}, {Ryan}, {Sevilla}, {Sonnett}, {Stevenson}, {Thompson}, {Wheelock},
  {Wiemer}, {Wittman}, {Wright}, \& {Yan}}]{Mainzer14}
{Mainzer}, A., {Bauer}, J., {Cutri}, R.~M., {et~al.} 2014, \apj, 792, 30

\bibitem[{{Maksym} {et~al.}(2013){Maksym}, {Ulmer}, {Eracleous}, {Guennou}, \&
  {Ho}}]{Maksym13}
{Maksym}, W.~P., {Ulmer}, M.~P., {Eracleous}, M.~C., {Guennou}, L., \& {Ho},
  L.~C. 2013, \mnras, 435, 1904

\bibitem[{{Margutti} {et~al.}(2017){Margutti}, {Metzger}, {Chornock},
  {Milisavljevic}, {Berger}, {Blanchard}, {Guidorzi}, {Migliori}, {Kamble},
  {Lunnan}, {Nicholl}, {Coppejans}, {Dall'Osso}, {Drout}, {Perna}, \&
  {Sbarufatti}}]{Margutti17}
{Margutti}, R., {Metzger}, B.~D., {Chornock}, R., {et~al.} 2017, \apj, 836, 25

\bibitem[{{Martin} {et~al.}(2005){Martin}, {Fanson}, {Schiminovich},
  {Morrissey}, {Friedman}, {Barlow}, {Conrow}, {Grange}, {Jelinsky},
  {Milliard}, {Siegmund}, {Bianchi}, {Byun}, {Donas}, {Forster}, {Heckman},
  {Lee}, {Madore}, {Malina}, {Neff}, {Rich}, {Small}, {Surber}, {Szalay},
  {Welsh}, \& {Wyder}}]{Martin05}
{Martin}, D.~C., {Fanson}, J., {Schiminovich}, D., {et~al.} 2005, \apjl, 619,
  L1

\bibitem[{{Merloni} {et~al.}(2015){Merloni}, {Dwelly}, {Salvato},
  {Georgakakis}, {Greiner}, {Krumpe}, {Nandra}, {Ponti}, \& {Rau}}]{Merloni15}
{Merloni}, A., {Dwelly}, T., {Salvato}, M., {et~al.} 2015, \mnras, 452, 69

\bibitem[{{Metzger} \& {Stone}(2016)}]{Metzger16}
{Metzger}, B.~D., \& {Stone}, N.~C. 2016, \mnras, 461, 948

\bibitem[{{Metzger} \& {Stone}(2017)}]{Metzger17}
---. 2017, \apj, 844, 75

\bibitem[{{Meyer} \& {Meyer-Hofmeister}(1981)}]{Meyer81}
{Meyer}, F., \& {Meyer-Hofmeister}, E. 1981, \aap, 104, L10

\bibitem[{{Mineshige} \& {Shields}(1990)}]{Mineshige90}
{Mineshige}, S., \& {Shields}, G.~A. 1990, \apj, 351, 47

\bibitem[{{Oke}(1974)}]{oke74}
{Oke}, J.~B. 1974, \apjs, 27, 21

\bibitem[{{Pasham} {et~al.}(2017){Pasham}, {Cenko}, {Sadowski}, {Guillochon},
  {Stone}, {van Velzen}, \& {Cannizzo}}]{Pasham17}
{Pasham}, D.~R., {Cenko}, S.~B., {Sadowski}, A., {et~al.} 2017, \apjl, 837, L30

\bibitem[{{Piran} {et~al.}(2015){Piran}, {Svirski}, {Krolik}, {Cheng}, \&
  {Shiokawa}}]{Piran15b}
{Piran}, T., {Svirski}, G., {Krolik}, J., {Cheng}, R.~M., \& {Shiokawa}, H.
  2015, \apj, 806, 164

\bibitem[{{Rees}(1988)}]{Rees88}
{Rees}, M.~J. 1988, \nat, 333, 523

\bibitem[{{Reis} {et~al.}(2012){Reis}, {Soares-Santos}, {Annis}, {Dodelson},
  {Hao}, {Johnston}, {Kubo}, {Lin}, {Seo}, \& {Simet}}]{Reis12}
{Reis}, R.~R.~R., {Soares-Santos}, M., {Annis}, J., {et~al.} 2012, \apj, 747,
  59

\bibitem[{{Reynolds}(2014)}]{Reynolds14}
{Reynolds}, C.~S. 2014, \ssr, 183, 277

\bibitem[{{Risaliti} {et~al.}(2013){Risaliti}, {Harrison}, {Madsen}, {Walton},
  {Boggs}, {Christensen}, {Craig}, {Grefenstette}, {Hailey}, {Nardini},
  {Stern}, \& {Zhang}}]{Risaliti13}
{Risaliti}, G., {Harrison}, F.~A., {Madsen}, K.~K., {et~al.} 2013, \nat, 494,
  449

\bibitem[{{Ruan} {et~al.}(2016){Ruan}, {Anderson}, {Cales}, {Eracleous},
  {Green}, {Morganson}, {Runnoe}, {Shen}, {Wilkinson}, {Blanton}, {Dwelly},
  {Georgakakis}, {Greene}, {LaMassa}, {Merloni}, \& {Schneider}}]{Ruan16}
{Ruan}, J.~J., {Anderson}, S.~F., {Cales}, S.~L., {et~al.} 2016, \apj, 826, 188

\bibitem[{{Saxton} {et~al.}(2016){Saxton}, {Perets}, \& {Baskin}}]{Saxton16}
{Saxton}, C.~J., {Perets}, H.~B., \& {Baskin}, A. 2016, ArXiv e-prints,
  arXiv:1612.08093

\bibitem[{{Saxton} {et~al.}(2017){Saxton}, {Read}, {Komossa}, {Lira},
  {Alexander}, \& {Wieringa}}]{Saxton17}
{Saxton}, R.~D., {Read}, A.~M., {Komossa}, S., {et~al.} 2017, \aap, 598, A29

\bibitem[{{Schlafly} \& {Finkbeiner}(2011)}]{Schlafly11}
{Schlafly}, E.~F., \& {Finkbeiner}, D.~P. 2011, \apj, 737, 103

\bibitem[{{Schmidt}(1968)}]{Schmidt68}
{Schmidt}, M. 1968, \apj, 151, 393

\bibitem[{{Shankar} {et~al.}(2004){Shankar}, {Salucci}, {Granato}, {De Zotti},
  \& {Danese}}]{Shankar04}
{Shankar}, F., {Salucci}, P., {Granato}, G.~L., {De Zotti}, G., \& {Danese}, L.
  2004, \mnras, 354, 1020

\bibitem[{{Shappee} {et~al.}(2014){Shappee}, {Prieto}, {Grupe}, {Kochanek},
  {Stanek}, {De Rosa}, {Mathur}, {Zu}, {Peterson}, {Pogge}, {Komossa}, {Im},
  {Jencson}, {Holoien}, {Basu}, {Beacom}, {Szczygie{\l}}, {Brimacombe},
  {Adams}, {Campillay}, {Choi}, {Contreras}, {Dietrich}, {Dubberley},
  {Elphick}, {Foale}, {Giustini}, {Gonzalez}, {Hawkins}, {Howell}, {Hsiao},
  {Koss}, {Leighly}, {Morrell}, {Mudd}, {Mullins}, {Nugent}, {Parrent},
  {Phillips}, {Pojmanski}, {Rosing}, {Ross}, {Sand}, {Terndrup}, {Valenti},
  {Walker}, \& {Yoon}}]{Shappee14}
{Shappee}, B.~J., {Prieto}, J.~L., {Grupe}, D., {et~al.} 2014, \apj, 788, 48

\bibitem[{{Siemiginowska} \& {Elvis}(1997)}]{Siemiginowsk97}
{Siemiginowska}, A., \& {Elvis}, M. 1997, \apjl, 482, L9

\bibitem[{{Skrutskie} {et~al.}(1997){Skrutskie}, {Schneider}, {Stiening},
  {Strom}, {Weinberg}, {Beichman}, {Chester}, {Cutri}, {Lonsdale}, {Elias},
  {Elston}, {Capps}, {Carpenter}, {Huchra}, {Liebert}, {Monet}, {Price}, \&
  {Seitzer}}]{Skrutskie97}
{Skrutskie}, M.~F., {Schneider}, S.~E., {Stiening}, R., {et~al.} 1997, in
  Astrophysics and Space Science Library, Vol. 210, The Impact of Large Scale
  Near-IR Sky Surveys, ed. F.~{Garzon}, N.~{Epchtein}, A.~{Omont}, B.~{Burton},
  \& P.~{Persi}, 25

\bibitem[{{Smak}(1983)}]{Smak83}
{Smak}, J. 1983, \apj, 272, 234

\bibitem[{{Stefanon} {et~al.}(2017){Stefanon}, {Yan}, {Mobasher}, {Barro},
  {Donley}, {Fontana}, {Hemmati}, {Koekemoer}, {Lee}, {Lee}, {Nayyeri}, {Peth},
  {Pforr}, {Salvato}, {Wiklind}, {Wuyts}, {Ashby}, {Castellano}, {Conselice},
  {Cooper}, {Cooray}, {Dolch}, {Ferguson}, {Galametz}, {Giavalisco}, {Guo},
  {Willner}, {Dickinson}, {Faber}, {Fazio}, {Gardner}, {Gawiser}, {Grazian},
  {Grogin}, {Kocevski}, {Koo}, {Lee}, {Lucas}, {McGrath}, {Nandra}, {Newman},
  \& {van der Wel}}]{Stefanon17}
{Stefanon}, M., {Yan}, H., {Mobasher}, B., {et~al.} 2017, \apjs, 229, 32

\bibitem[{{Stern} {et~al.}(2012){Stern}, {Assef}, {Benford}, {Blain}, {Cutri},
  {Dey}, {Eisenhardt}, {Griffith}, {Jarrett}, {Lake}, {Masci}, {Petty},
  {Stanford}, {Tsai}, {Wright}, {Yan}, {Harrison}, \& {Madsen}}]{Stern12}
{Stern}, D., {Assef}, R.~J., {Benford}, D.~J., {et~al.} 2012, \apj, 753, 30

\bibitem[{{Stone} \& {Metzger}(2016)}]{Stone16b}
{Stone}, N.~C., \& {Metzger}, B.~D. 2016, \mnras, 455, 859

\bibitem[{{Stone} \& {van Velzen}(2016)}]{StonevanVelzen16}
{Stone}, N.~C., \& {van Velzen}, S. 2016, \apjl, 825, L14

\bibitem[{{Storchi-Bergmann} {et~al.}(1993){Storchi-Bergmann}, {Baldwin}, \&
  {Wilson}}]{Storchi-Berg93}
{Storchi-Bergmann}, T., {Baldwin}, J.~A., \& {Wilson}, A.~S. 1993, \apjl, 410,
  L11

\bibitem[{{Stoughton} {et~al.}(2002){Stoughton}, {Lupton}, {Bernardi},
  {Blanton}, {Burles}, {Castander}, {Connolly}, {Eisenstein}, {Frieman},
  {Hennessy}, {Hindsley}, {Ivezi{\'c}}, {Kent}, {Kunszt}, {Lee}, {Meiksin},
  {Munn}, {Newberg}, {Nichol}, {Nicinski}, {Pier}, {Richards}, {Richmond},
  {Schlegel}, {Smith}, {Strauss}, {SubbaRao}, {Szalay}, {Thakar}, {Tucker},
  {Vanden Berk}, {Yanny}, {Adelman}, {Anderson}, {Anderson}, {Annis},
  {Bahcall}, {Bakken}, {Bartelmann}, {Bastian}, {Bauer}, {Berman},
  {B{\"o}hringer}, {Boroski}, {Bracker}, {Briegel}, {Briggs}, {Brinkmann},
  {Brunner}, {Carey}, {Carr}, {Chen}, {Christian}, {Colestock}, {Crocker},
  {Csabai}, {Czarapata}, {Dalcanton}, {Davidsen}, {Davis}, {Dehnen},
  {Dodelson}, {Doi}, {Dombeck}, {Donahue}, {Ellman}, {Elms}, {Evans}, {Eyer},
  {Fan}, {Federwitz}, {Friedman}, {Fukugita}, {Gal}, {Gillespie}, {Glazebrook},
  {Gray}, {Grebel}, {Greenawalt}, {Greene}, {Gunn}, {de Haas}, {Haiman},
  {Haldeman}, {Hall}, {Hamabe}, {Hansen}, {Harris}, {Harris}, {Harvanek},
  {Hawley}, {Hayes}, {Heckman}, {Helmi}, {Henden}, {Hogan}, {Hogg}, {Holmgren},
  {Holtzman}, {Huang}, {Hull}, {Ichikawa}, {Ichikawa}, {Johnston}, {Kauffmann},
  {Kim}, {Kimball}, {Kinney}, {Klaene}, {Kleinman}, {Klypin}, {Knapp},
  {Korienek}, {Krolik}, {Kron}, {Krzesi{\'n}ski}, {Lamb}, {Leger},
  {Limmongkol}, {Lindenmeyer}, {Long}, {Loomis}, {Loveday}, {MacKinnon},
  {Mannery}, {Mantsch}, {Margon}, {McGehee}, {McKay}, {McLean}, {Menou},
  {Merelli}, {Mo}, {Monet}, {Nakamura}, {Narayanan}, {Nash}, {Neilsen},
  {Newman}, {Nitta}, {Odenkirchen}, {Okada}, {Okamura}, {Ostriker}, {Owen},
  {Pauls}, {Peoples}, {Peterson}, {Petravick}, {Pope}, {Pordes}, {Postman},
  {Prosapio}, {Quinn}, {Rechenmacher}, {Rivetta}, {Rix}, {Rockosi}, {Rosner},
  {Ruthmansdorfer}, {Sandford}, {Schneider}, {Scranton}, {Sekiguchi}, {Sergey},
  {Sheth}, {Shimasaku}, {Smee}, {Snedden}, {Stebbins}, {Stubbs}, {Szapudi},
  {Szkody}, {Szokoly}, {Tabachnik}, {Tsvetanov}, {Uomoto}, {Vogeley}, {Voges},
  {Waddell}, {Walterbos}, {Wang}, {Watanabe}, {Weinberg}, {White}, {White},
  {Wilhite}, {Wolfe}, {Yasuda}, {York}, {Zehavi}, \& {Zheng}}]{stoughton02}
{Stoughton}, C., {Lupton}, R.~H., {Bernardi}, M., {et~al.} 2002, \aj, 123, 485

\bibitem[{{Strubbe} \& {Quataert}(2011)}]{Strubbe11}
{Strubbe}, L.~E., \& {Quataert}, E. 2011, \mnras, 415, 168

\bibitem[{{Tanaka}(2013)}]{Tanaka13}
{Tanaka}, T.~L. 2013, \mnras, 434, 2275

\bibitem[{{Taylor} {et~al.}(2010){Taylor}, {Franx}, {Brinchmann}, {van der
  Wel}, \& {van Dokkum}}]{Taylor10}
{Taylor}, E.~N., {Franx}, M., {Brinchmann}, J., {van der Wel}, A., \& {van
  Dokkum}, P.~G. 2010, \apj, 722, 1

\bibitem[{{van Paradijs}(1996)}]{van-Paradijs96}
{van Paradijs}, J. 1996, \apjl, 464, L139

\bibitem[{{van Velzen} \& {Farrar}(2014)}]{vanVelzen14}
{van Velzen}, S., \& {Farrar}, G.~R. 2014, \apj, 792, 53

\bibitem[{van Velzen {et~al.}(2011)van Velzen, K\"ording, \&
  Falcke}]{vanVelzen11}
van Velzen, S., K\"ording, E., \& Falcke, H. 2011, \mnras, 417, L51

\bibitem[{{van Velzen} {et~al.}(2016{\natexlab{a}}){van Velzen}, {Mendez},
  {Krolik}, \& {Gorjian}}]{vanVelzen16b}
{van Velzen}, S., {Mendez}, A.~J., {Krolik}, J.~H., \& {Gorjian}, V.
  2016{\natexlab{a}}, \apj, 829, 19

\bibitem[{{van Velzen} {et~al.}(2011){van Velzen}, {Farrar}, {Gezari},
  {Morrell}, {Zaritsky}, {{\"O}stman}, {Smith}, {Gelfand}, \&
  {Drake}}]{vanVelzen10}
{van Velzen}, S., {Farrar}, G.~R., {Gezari}, S., {et~al.} 2011, \apj, 741, 73

\bibitem[{{van Velzen} {et~al.}(2016{\natexlab{b}}){van Velzen}, {Anderson},
  {Stone}, {Fraser}, {Wevers}, {Metzger}, {Jonker}, {van der Horst}, {Staley},
  {Mendez}, {Miller-Jones}, {Hodgkin}, {Campbell}, \& {Fender}}]{vanVelzen16}
{van Velzen}, S., {Anderson}, G.~E., {Stone}, N.~C., {et~al.}
  2016{\natexlab{b}}, Science, 351, 62

\bibitem[{{Wang} \& {Merritt}(2004)}]{Wang04}
{Wang}, J., \& {Merritt}, D. 2004, \apj, 600, 149

\bibitem[{{Wevers} {et~al.}(2017){Wevers}, {van Velzen}, {Jonker}, {Stone},
  {Hung}, {Onori}, {Gezari}, \& {Blagorodnova}}]{Wevers17}
{Wevers}, T., {van Velzen}, S., {Jonker}, P.~G., {et~al.} 2017, \mnras, 471,
  1694

\bibitem[{{Wright} {et~al.}(2010){Wright}, {Eisenhardt}, {Mainzer}, {Ressler},
  {Cutri}, {Jarrett}, {Kirkpatrick}, {Padgett}, {McMillan}, {Skrutskie},
  {Stanford}, {Cohen}, {Walker}, {Mather}, {Leisawitz}, {Gautier}, {McLean},
  {Benford}, {Lonsdale}, {Blain}, {Mendez}, {Irace}, {Duval}, {Liu}, {Royer},
  {Heinrichsen}, {Howard}, {Shannon}, {Kendall}, {Walsh}, {Larsen}, {Cardon},
  {Schick}, {Schwalm}, {Abid}, {Fabinsky}, {Naes}, \& {Tsai}}]{Wright10}
{Wright}, E.~L., {Eisenhardt}, P.~R.~M., {Mainzer}, A.~K., {et~al.} 2010, \aj,
  140, 1868

\bibitem[{{Wyrzykowski} {et~al.}(2017){Wyrzykowski}, {Zieli{\'n}ski},
  {Kostrzewa-Rutkowska}, {Hamanowicz}, {Jonker}, {Arcavi}, {Guillochon},
  {Brown}, {Koz{\l}owski}, {Udalski}, {Szyma{\'n}ski}, {Soszy{\'n}ski},
  {Poleski}, {Pietrukowicz}, {Skowron}, {Mr{\'o}z}, {Ulaczyk}, {Pawlak},
  {Rybicki}, {Greiner}, {Kr{\"u}hler}, {Bolmer}, {Smartt}, {Maguire}, \&
  {Smith}}]{Wyrzykowski17}
{Wyrzykowski}, {\L}., {Zieli{\'n}ski}, M., {Kostrzewa-Rutkowska}, Z., {et~al.}
  2017, \mnras, 465, L114

\bibitem[{{York} {et~al.}(2000){York}, {Adelman}, {Anderson}, {Anderson},
  {Annis}, {Bahcall}, {Bakken}, {Barkhouser}, {Bastian}, {Berman}, {Boroski},
  {Bracker}, {Briegel}, {Briggs}, {Brinkmann}, {Brunner}, {Burles}, {Carey},
  {Carr}, {Castander}, {Chen}, {Colestock}, {Connolly}, {Crocker}, {Csabai},
  {Czarapata}, {Davis}, {Doi}, {Dombeck}, {Eisenstein}, {Ellman}, {Elms},
  {Evans}, {Fan}, {Federwitz}, {Fiscelli}, {Friedman}, {Frieman}, {Fukugita},
  {Gillespie}, {Gunn}, {Gurbani}, {de Haas}, {Haldeman}, {Harris}, {Hayes},
  {Heckman}, {Hennessy}, {Hindsley}, {Holm}, {Holmgren}, {Huang}, {Hull},
  {Husby}, {Ichikawa}, {Ichikawa}, {Ivezi{\'c}}, {Kent}, {Kim}, {Kinney},
  {Klaene}, {Kleinman}, {Kleinman}, {Knapp}, {Korienek}, {Kron}, {Kunszt},
  {Lamb}, {Lee}, {Leger}, {Limmongkol}, {Lindenmeyer}, {Long}, {Loomis},
  {Loveday}, {Lucinio}, {Lupton}, {MacKinnon}, {Mannery}, {Mantsch}, {Margon},
  {McGehee}, {McKay}, {Meiksin}, {Merelli}, {Monet}, {Munn}, {Narayanan},
  {Nash}, {Neilsen}, {Neswold}, {Newberg}, {Nichol}, {Nicinski}, {Nonino},
  {Okada}, {Okamura}, {Ostriker}, {Owen}, {Pauls}, {Peoples}, {Peterson},
  {Petravick}, {Pier}, {Pope}, {Pordes}, {Prosapio}, {Rechenmacher}, {Quinn},
  {Richards}, {Richmond}, {Rivetta}, {Rockosi}, {Ruthmansdorfer}, {Sandford},
  {Schlegel}, {Schneider}, {Sekiguchi}, {Sergey}, {Shimasaku}, {Siegmund},
  {Smee}, {Smith}, {Snedden}, {Stone}, {Stoughton}, {Strauss}, {Stubbs},
  {SubbaRao}, {Szalay}, {Szapudi}, {Szokoly}, {Thakar}, {Tremonti}, {Tucker},
  {Uomoto}, {Vanden Berk}, {Vogeley}, {Waddell}, {Wang}, {Watanabe},
  {Weinberg}, {Yanny}, \& {Yasuda}}]{york02}
{York}, D.~G., {Adelman}, J., {Anderson}, Jr., J.~E., {et~al.} 2000, \aj, 120,
  1579

\bibitem[{{Zabludoff} {et~al.}(1996){Zabludoff}, {Zaritsky}, {Lin}, {Tucker},
  {Hashimoto}, {Shectman}, {Oemler}, \& {Kirshner}}]{Zabludoff96}
{Zabludoff}, A.~I., {Zaritsky}, D., {Lin}, H., {et~al.} 1996, \apj, 466, 104

\bibitem[{{Zauderer} {et~al.}(2011){Zauderer}, {Berger}, {Soderberg}, {Loeb},
  {Narayan}, {Frail}, {Petitpas}, {Brunthaler}, {Chornock}, {Carpenter},
  {Pooley}, {Mooley}, {Kulkarni}, {Margutti}, {Fox}, {Nakar}, {Patel},
  {Volgenau}, {Culverhouse}, {Bietenholz}, {Rupen}, {Max-Moerbeck}, {Readhead},
  {Richards}, {Shepherd}, {Storm}, \& {Hull}}]{Zauderer11}
{Zauderer}, B.~A., {Berger}, E., {Soderberg}, A.~M., {et~al.} 2011, \nat, 476,
  425

\end{thebibliography}


\clearpage

\appendix

%
\section{Synthetic Galaxy Sample}
%
Our synthetic galaxy sample, discussed in Sec.~\ref{sec:syngal}, is available at the journal website. In Table~\ref{tab:syngal} we list the columns of this catalog.

\begin{deluxetable*}{l l l c c c c c c}
\tablewidth{0pt}
\tablecolumns{3}
\tablecaption{Columns of the synthetic galaxy catalog }
\tablehead{column name & unit & comments}
\startdata
\verb z 										& 							& Redshift  \\
\verb ra 									& deg 					& Right Ascension (of original galaxy) \\
\verb dec 								& deg 					& Declination (of original galaxy) \\
\verb mass 							& $M_{\odot}$ & Total galaxy mass, from NYU-VAGC, based on {\it ugrizJHK} photometry \\
\verb B300 							& yr$^{-1}$ 		& Specific SFR over the past 300 Myr, from NYU-VAGC, based on {\it ugrizJHK} photometry \\
\verb B1000 							& yr$^{-1}$ 		& Specific SFR over the past Gyr, from NYU-VAGC, based on {\it ugrizJHK} photometry \\
\verb sSFR	  							& yr$^{-1}$ 		& Specific SFR, from the MPA-JHU catalog (their \verb specsfr_fib_p50  column) \\
\verb BT  								&  							& Bulge-to-total ratio, based on Lackner \& Gunn (2012) measurements in the $r$ band \\
\verb r50_kpc  						&  kpc 					& Effective radius based on Sersic fit from NYU-VAGC  \\
\verb sersic_n  					&  							& Sersic index, from NYU-VAGC  \\
\verb sigma  							&  km s$^{-1}$	& Velocity dispersion as estimated using the virial theorem (Eq.~\ref{eq:sigma}) \\
\verb sigma_SDSS  			&  km s$^{-1}$	& Velocity dispersion from SDSS pipeline (as reported in the NYU-VAGC) \\
\verb sigma_SDSS_err  	&  km s$^{-1}$	& Uncertainty on \verb sigma_SDSS     \\
\verb MBH_sigma  			&  $M_{\odot}$	& Black hole mass as estimated from the velocity dispersion (Eq.~\ref{eq:gulte_sigma}) \\
\verb MBH_bulge  				&  $M_{\odot}$	& Black hole mass as estimated from the bulge mass (Eq.~\ref{eq:gulte_bulge}) \\
\verb m_r  								& AB mag 			& Apparent magnitude in the $r$ band \\
\verb M_r  								& AB mag 			& Absolute magnitude in the $r$ band (k-corrected) \\
\verb m_g  							& AB mag 			& Apparent magnitude in the $g$ band \\
\verb M_g    							& AB mag 			& Absolute magnitude in the $g$ band (k-corrected) \\
\enddata
\label{tab:syngal}
\end{deluxetable*}

\end{document}